\documentclass[10pt,twocolumn,letterpaper]{article}

\usepackage{iccv}              
\usepackage{graphicx}
\usepackage{amsmath}
\usepackage{amssymb}
\usepackage{adjustbox}
\usepackage{array}
\newcolumntype{C}[1]{>{\centering\arraybackslash}m{#1}}
\usepackage{booktabs}
\usepackage{color, colortbl}
\usepackage{amsmath,amssymb,amsfonts}
\usepackage{algorithmic}
\usepackage{graphicx}
\usepackage{textcomp}
\usepackage{xcolor}
\usepackage{makecell}
\usepackage{pifont}       
\usepackage{bbding}       
\usepackage{fontawesome}  
\usepackage{svg} 
\usepackage{multirow}
\usepackage{booktabs} 
\usepackage{threeparttable}
\usepackage{dsfont}
\usepackage{xcolor}
\newcommand{\cmark}{\ding{51}}
\newcommand{\xmark}{\ding{55}}

\newcommand\figcaption{\def\@captype{figure}\caption}
\newcommand\tabcaption{\def\@captype{table}\caption}

\newcommand\blfootnote[1]{%
  \begingroup
  \renewcommand\thefootnote{}\footnote{#1}%
  \addtocounter{footnote}{-1}%
  \endgroup
}

\definecolor{darkgreen}{rgb}{0.0, 0.8, 0.0}
\definecolor{backred}{RGB}{255, 190, 190}
\definecolor{backblue}{RGB}{210, 230, 250}
\definecolor{backgreen}{RGB}{137, 215, 188}
\definecolor{backyellow}{RGB}{251, 218, 195}
\definecolor{cvprblue}{rgb}{0.21,0.49,0.74}
\usepackage[pagebackref,breaklinks,colorlinks,allcolors=cvprblue]{hyperref}

\def\paperID{9157} 
\def\confName{ICCV}
\def\confYear{2025}

\title{\texttt{SAMora}: Enhancing SAM through  Hierarchical Self-Supervised \\ Pre-Training for Medical Images}

\author{
Shuhang Chen$^{1\dagger}$ \quad Hangjie Yuan$^{1\dagger}$ \quad Pengwei Liu$^{1}$ \quad Hanxue Gu$^{2}$ \\
\quad Tao Feng$^{3}$  \quad Dong Ni$^{1\star}$ \vspace{.2cm} \\
{
$^{1}$Zhejiang University \quad
$^{2}$Duke  University \quad
$^{3}$Tsinghua University 
}
}

\begin{document}
\maketitle
\begin{abstract}
The Segment Anything Model (SAM) has demonstrated significant potential in medical image segmentation. Yet, its performance is limited when only a small amount of labeled data is available, while there is abundant valuable yet often overlooked hierarchical information in medical data. To address this limitation, we draw inspiration from self-supervised learning and propose SAMora, an innovative framework that captures hierarchical medical knowledge by applying complementary self-supervised learning objectives at the image, patch, and pixel levels. To fully exploit the complementarity of hierarchical knowledge within LoRAs, we introduce HL-Attn, a hierarchical fusion module that integrates multi-scale features while maintaining their distinct characteristics. SAMora is compatible with various SAM variants, including SAM2, SAMed, and H-SAM. 
Experimental results on the Synapse, LA, and PROMISE12 datasets demonstrate that SAMora outperforms existing SAM variants. It achieves state-of-the-art performance in both few-shot and fully supervised settings while reducing fine-tuning epochs by 90\%. 
The code is available at \href{https://github.com/ShChen233/SAMora}{\textcolor{red}{\texttt{https://github.com/ShChen233/SAMora}}}
.
\end{abstract}    
\blfootnote{$^{\dagger}$ These authors contributed equally to this work.}
\blfootnote{$^{\star}$ Corresponding author: dni@zju.edu.cn.}
\section{Introduction}
\label{sec:intro}
The Segmentation Anything Model (SAM)~\cite{sam} stands as one of the most versatile and comprehensive foundational models in the field of image segmentation, gaining widespread recognition for its adaptability and high performance across a broad range of applications. Its effectiveness has been demonstrated in diverse domains, including satellite imagery analysis~\cite{sate} and autonomous driving~\cite{autodriving}, where robust and accurate segmentation is critical.

However, SAM's performance is notably less impressive when directly applied to medical images~\cite{samad, samunet}. 
The inherent complexity of medical imaging, coupled with the model's reliance on prompts, poses significant challenges for its direct application in this domain. 
Additionally, the scarcity of training medical images—due to legal restrictions and annotation difficulty  further exacerbates these challenges, leading to suboptimal segmentation results in medical imaging tasks.
To overcome this limitation, researchers have turned to prompt-free fine-tuning techniques, leveraging the available labeled data to adapt SAM more effectively to the medical domain. Efforts such as SAMed~\cite{samed} and H-SAM~\cite{hsam} have demonstrated that fine-tuning SAM with domain-specific data significantly enhances its ability to capture the unique patterns inherent in medical images.
\begin{figure}[t]
    \centering
    \includegraphics[width=1\linewidth]{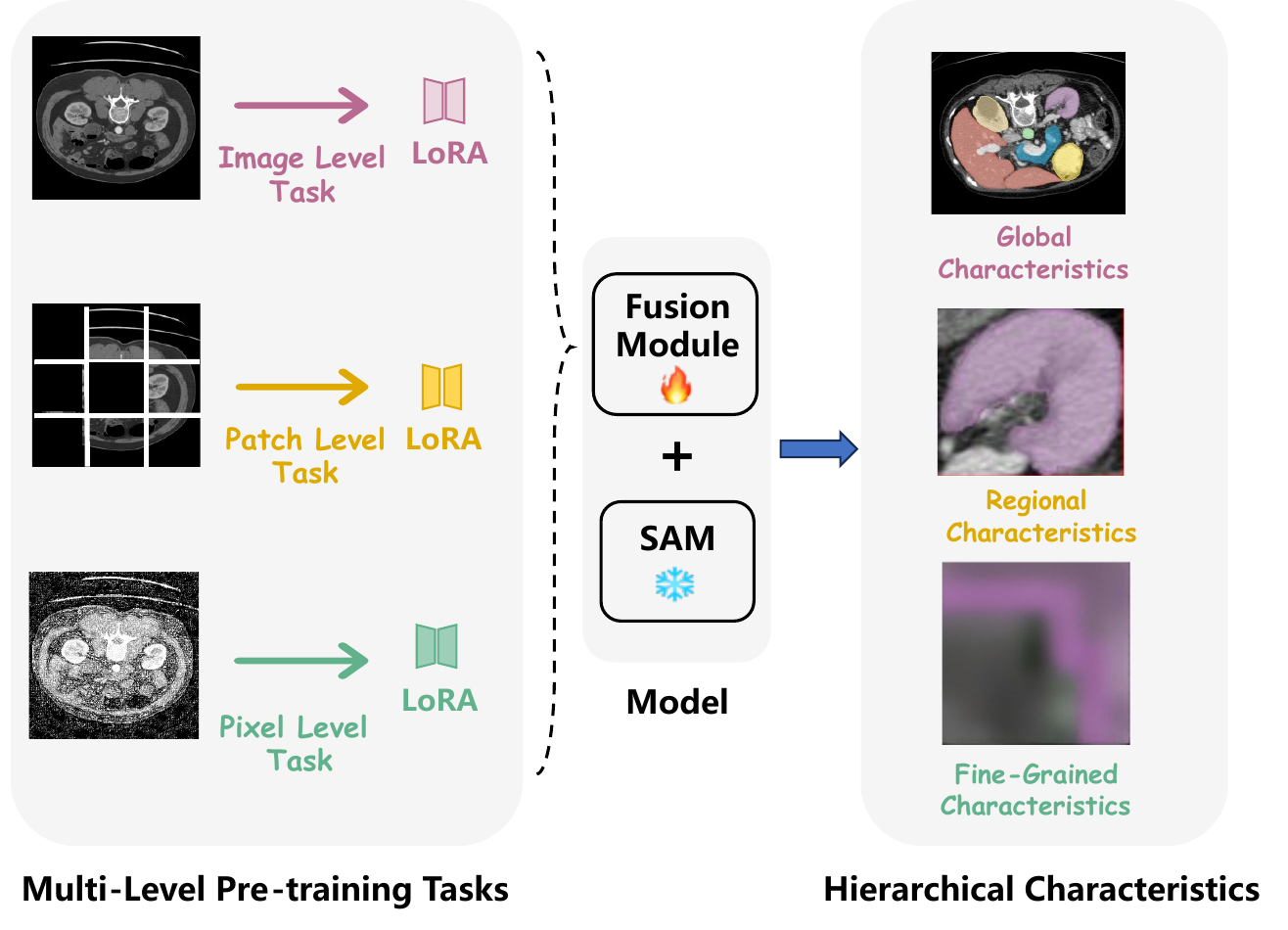}
    \caption{
    \textbf{The Hierarchical Characteristics of Multi-Level Pre-training Tasks on Medical Images.} The abundant hierarchical characteristics inherent in vast amounts of unlabeled data, when effectively fused, can significantly enhance the segmentation performance of SAM.
            }
    \label{fig:enter-label}
    \vspace {-1.5em}
\end{figure}
These advancements have paved the way for more accurate and reliable medical image segmentation; however, they often \textit{overlook the vast amounts of unlabeled data available}, leaving a wealth of potentially valuable information untapped.

Recent advancements in self-supervised learning(SSL), such as masked autoencoder (MAE)~\cite{mae} and contrastive learning~\cite{cl}, have gained substantial attention for their ability to leverage unlabeled data without the need for manual annotation. Studies have demonstrated that models trained using self-supervised learning techniques can even surpass the performance of those trained exclusively on labeled data~\cite{sam_finetune}. Given the abundance of unlabeled data in the medical imaging domain, this approach presents a significant opportunity to enhance SAM's adaptability to the complex and diverse characteristics of medical images. These promising developments have inspired us to explore the integration of self-supervised learning into fine-tuning the SAM model.

Furthermore, we observed that medical images exhibit hierarchical and multi-scale structures~\cite{hifuse,hierarchical1}, with each scale providing unique features that are critical for accurate diagnosis. 
As depicted in Fig.~\ref{fig:enter-label}, The hierarchical characteristics of the multi-level pre-training
tasks on the medical image show that the image-level patterns capture global characteristics, which is a broad overview of the anatomical region 
and essential for identifying general structures and context. 
The patch-level patterns allow for a more detailed examination of specific regions called regional characteristics, highlighting finer anatomical features, while the pixel-level patterns offer fine-grained characteristics with the highest resolution, enabling the detection of subtle tissue variations. We hypothesize that \textit{these hierarchical patterns are complementary to each other and that combining them with self-supervised learning on unlabeled data can significantly improve segmentation performance.}

In response to these insights, we propose \textbf{SAMora}, a prompt-free fine-tuned SAM model that incorporates multiple LoRA experts, which leverage large amounts of unlabeled medical image data via two stages. 
The first stage focuses on enhancing SAM model with hierarchical medical knowledge, with the use of hierarchical self-supervised pre-training. 
Specifically, we pre-train LoRA experts for the SAM model in the image, patch, and pixel levels, which are used for follow-up fine-tuning.
Notably, the LoRA experts for image and patch levels are learned by distilling from continually pre-trained teacher models (\textit{i.e.}, SimCLRv2~\cite{simclrv2} and MAE~\cite{mae}) that enables more effective medical knowledge awareness.

The second stage involves fine-tuning with labeled data, during which we introduce a hierarchical attention mechanism, HL-Attn (Hierarchical LoRA Attention). HL-Attn leverages the hierarchical properties at each level, \textbf{first} progressively integrating features from lower to higher levels. By adaptively integrating and refining knowledge from multiple levels of medical representations, it effectively captures the hierarchical characteristics of each level. \textbf{Fig.~\textcolor{cvprblue}{6} in Appendix \textcolor{cvprblue}{B.4} shows that hierarchical fusion is key to the model's ability to handle complex medical imaging tasks.}

We perform comprehensive experiments on multi-organ segmentation datasets (\textit{i.e.}, Synapse, left atrial (LA), and PROMISE2012 datasets) in both fully supervised and few-shot settings, which demonstrates SAMora's consistent superiority over existing prompt-free SAM counterparts. 
Moreover, SAMora is compatible with different prompt-free SAM variants, such as SAMed~\cite{samed} (by default), H-SAM~\cite{hsam} where the decoder has been modified from SAMed. Notably, we also apply the SAMora on SAM2~\cite{sam2}, which is a novel segment anything model proposed recently, called \textbf{SAMora-2}.
Specifically, on the Synapse dataset, SAMora and its variant achieve a Mean Dice boost of \textbf{4.09\%} using only 10\% of the training data and a Mean Dice boost of 2.10\% when utilizing the whole dataset while consuming only \textbf{10\%} of the fine-tuning epochs compared with other SAM counterparts.

The core contributions can be summarized as follows:
\textbf{1)} We propose to integrate three hierarchical levels of self-supervised knowledge from unlabeled medical images to existing SAM variants by pre-training LoRA experts. 
\textbf{2)} We propose the HL-Attn module to adaptively fuse hierarchical medical knowledge, ensuring that the model fully exploits the information available across different scales. 
\textbf{3)} 
SAMora is comparable with different SAM variants and achieves state-of-the-art (SOTA) performance on the LA, PROMISE12, and Synapse datasets in fully supervised and few-shot settings with only 10\% of the fine-tuning epochs, highlighting its efficiency and effectiveness in medical image segmentation.
\section{Related Works}
\label{sec:relate}

\subsection{SAM and Related Fine-tuning Approaches} 
Recently, the Segment Anything Model (SAM) has gained significant attention as a robust foundation model for image segmentation~\cite{sam, efficientsam}. However, the adaptation of SAM from general-purpose image segmentation to more specialized domains, such as medical imaging, presents considerable challenges~\cite{sam-ft,huang2024segment,wei2023medsam,xu2024mc}. 
\begin{figure*}[t]
    \centering
    \includegraphics[width=1\linewidth]{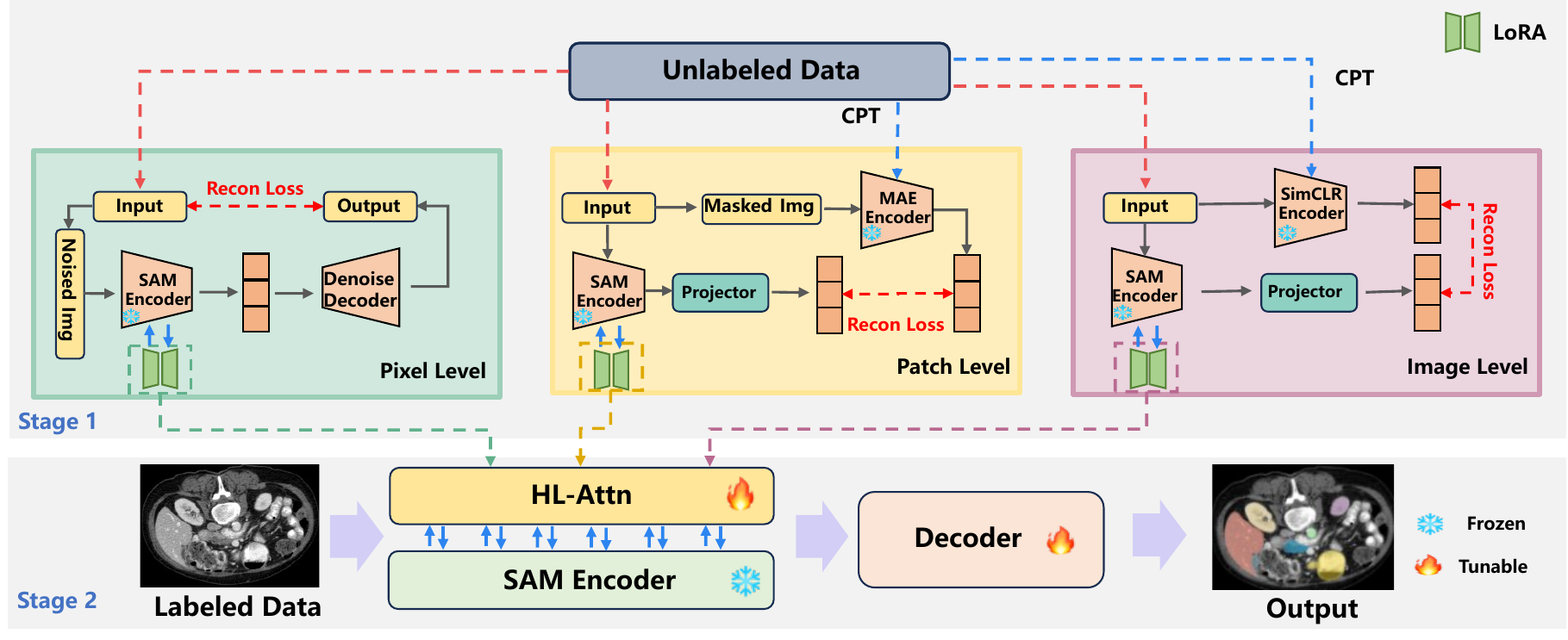}
    \caption{\textbf{The Overview of SAMora.} The training process of SAMora is divided into two stages. Stage 1 involves self-supervised pre-training using different LoRA experts across hierarchical levels. Each level employs a distinct self-supervised learning method: SimCLRv2 for the image level, MAE for the patch level, and denoising autoencoder for the pixel level. Continual Pre-Training (CPT) is applied to adapt the teacher models (SimCLRv2 and MAE)  to the medical imaging domain. Stage 2 focuses on fine-tuning with labeled data, where the SAM encoder and LoRA experts remain frozen, and only the HL-Attn and Decoder components are tuned. The projector is a trainable dimension-alignment module.}
    \label{fig:o1}
    \vspace{-2em}
\end{figure*}

Prompt-based SAM variants, which leverage user-defined prompts to guide the segmentation process, have shown promise in improving the model’s performance on specific tasks. For instance, SAM-Path~\cite{sampath} and MedSAM~\cite{ma2024segment13}, have demonstrated improved accuracy when providing manual prompts. However, the complexity of images and the need for clinical expertise make manual annotation impractical, while prompts introduce ambiguity due to varying interpretations and inconsistent capture of object structures.


Hence, prompt-free fine-tuning is proposed without the need for user-defined prompts. For instance, SAMed~\cite{samed} employs Low-Rank Adaptation (LoRA)~\cite{lora} to fine-tune the SAM model with labeled image data. Similarly, the Medical-SAM-Adapter~\cite{msa} leverages model adapters~\cite{adapter} to fine-tune SAM. However, given the limited amount of labeled data, which constrains the model's ability to learn all domain-specific information, we introduced unlabeled data by pre-training before fine-tuning.
\vspace{-1em}

\subsection{Self-Supervised Learning (SSL)}
Self-Supervised Learning (SSL) has emerged as a paradigm for utilizing unlabeled data in pre-training models due to its ability to learn robust feature representations without the need for manual annotations. Among the various SSL approaches, contrastive learning methods such as SimCLR~\cite{simclr}, MoCo~\cite{moco}, and InfoNCE~\cite{infonce} have been widely adopted for their effectiveness in distinguishing between similar and dissimilar samples by contrasting positive pairs with negative pairs, which is suited for capturing global relationships and broader patterns.

Additionally, reconstruction-based approaches have gained considerable attention for their ability to capture features at finer scales. masked autoencoder (MAE)~\cite{mae}, denoising autoencoders~\cite{denoising1,denoising2}, and other related techniques~\cite{I-JEPA,beit} such as I-JEPA focuses on reconstructing corrupted or missing parts of the data. MAE, for example, emphasizes the reconstruction of missing patches in an image, which is ideal for learning intermediate features at the patch level, where understanding localized structures is key. Unfortunately, few researchers have explored the combination of different SSL methods to fully leverage their unique characteristics for improving downstream task performance. 

\subsection{Multi-LoRA Fusion}
The primary goal of multi-LoRA fusion techniques is to enhance model performance by effectively combining the outputs of different LoRA experts~\cite{arrow,loraretriever}, each of which may be specialized for different tasks or datasets.

Linear Arithmetic Composition (LAC)~\cite{lac1,lac2,loraflow} involves a simple linear combination of the outputs from each LoRA, such as LoraHub~\cite{lorahub}. However, this approach often fails to preserve the unique characteristics of each LoRA block. On the other hand, tuning-based composition~\cite{mixofshow} has been developed specifically for the vision-and-language domain. However, this method notably limited the flexibility of LoRA fusion, primarily due to its reliance on manually-designed masks.

To address these limitations, the Mixture of LoRA Experts (MOLE)~\cite{mole} introduces a gating mechanism that dynamically adjusts the contribution of each LoRA block based on the input data. However, MOLE feeds different experts with the same or similar sampled subsets. \textit{In contrast, our method introduces different levels of image features—ranging from image-level to patch-level—into the model, preventing capturing duplicated information across various model components.}
\section{Method}
\label{sec:method}

\subsection{Model Overview}\label{AA}

We propose SAMora, which leverages self-supervised learning and explores hierarchical feature fusion as a potential enhancement, as shown in Fig.~\ref{fig:o1}. SAMora follows a two-stage process: 

\begin{figure}[t]
    \centering
    \includegraphics[width=1\linewidth]{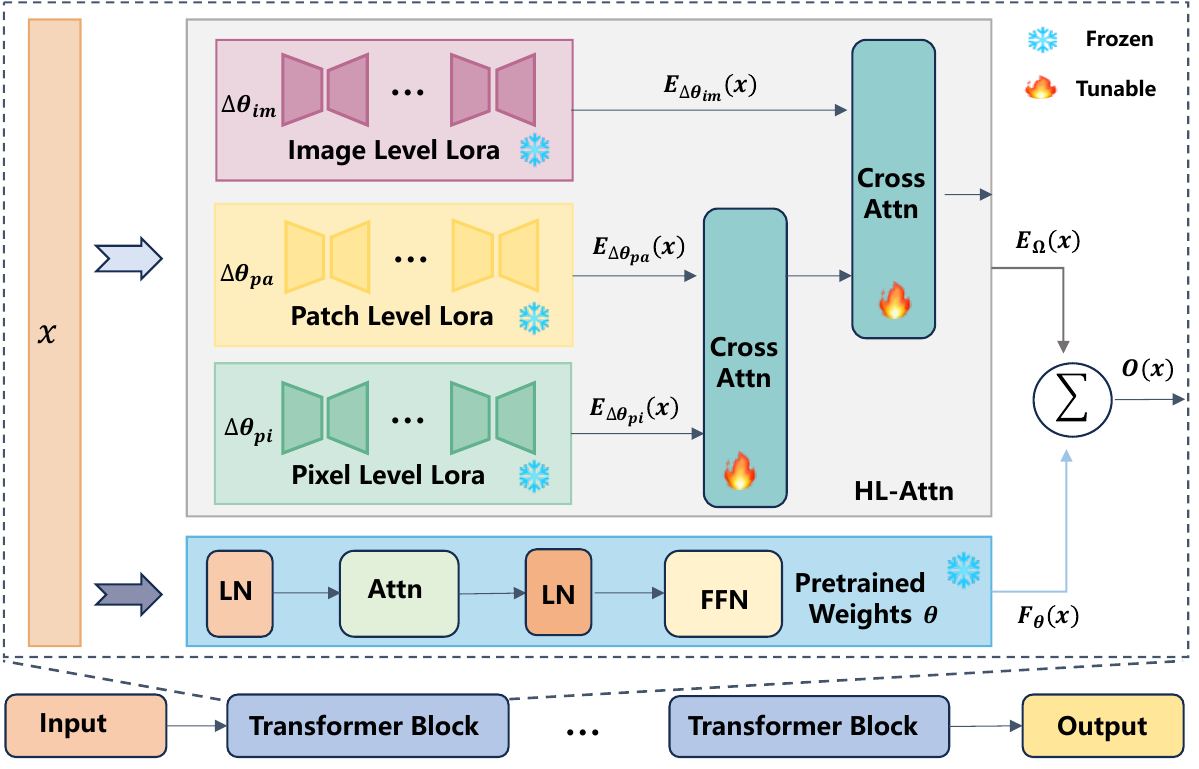}
    \vspace{-2em}
    \caption{\textbf{The Structure of HL-Attn.} Note that self-attention is not visualized in this figure.}
    \label{fig:hl-attn}
    \vspace{-1.5em}
\end{figure}
In stage 1, we pre-train the SAM model using LoRA, capitalizing on unlabeled medical image data abundance. Specifically, we employ multiple self-supervised learning methods: contrastive learning for image-level features, MAE for patch-level features, and denoising for pixel-level features. The LoRA experts are then passed into stage 2, where we fine-tune the SAM model using a small amount of labeled data. To effectively integrate the three LoRA experts, we propose HL-Attn (Hierarchical LoRA Attention), a hierarchical cross-attention mechanism that fuses features across different levels. This process involves sequentially merging features from the pixel level, patch level, and image level through multiple layers of cross-attention. During training, we freeze the SAM encoder and the pre-trained LoRA weights, allowing only the HL-Attn and Decoder weights to be updated. This approach ensures that the model benefits from the robust feature representations learned during pre-training while optimizing the integration of multi-level features for improved performance.

\subsection{SAMora: Self-Supervised Pre-Training Stage}


To capture the unique characteristics at three levels effectively, we employ different self-supervised learning methods tailored to each scale, as illustrated in Fig.~\ref{fig:hl-attn}.

\textbf{Image-Level Pre-Training: Contrastive Learning.} At the image level, the focus is on capturing global structures that are crucial for identifying broad anatomical features. Hence, we utilize SimCLRv2~\cite{simclrv2}.

As shown in Fig.~\ref{fig:o1}, the pretext task at the image level employs a teacher-student framework~\cite{ts} to distill knowledge from the teacher model into the student model. In this setup, the SimCLRv2 encoder functions as the teacher network, transferring its learned representations, while the SAM encoder, augmented with LoRA, serves as the student network that receives and integrates this distilled knowledge. Specifically, the teacher network is initialized using ResNet50 (2X+SK)~\cite{resnet} weights.  

Since these weights were originally trained on the ImageNet dataset, their performance in the medical domain is limited. To address this, we further \textbf{continual pre-train}~\cite{cpt2,cpt1} the SimCLRv2 weights on a dataset of 100,000 unlabeled medical images to better align the model's representations.

Given a mini-batch of augmented examples, the contrastive loss between a pair of positive examples $i,j$ (which are augmentations of the same image) is followed by SimCLRv2.

After fine-tuning, we freeze the weights of the SimCLRv2 model. During training, the SAM encoder remains frozen, with only the corresponding LoRA weights being updated. Additionally, following the approach used in EfficientSAM~\cite{efficientsam}, we employ a reconstruction loss to optimize.

\noindent \textbf{Patch-Level Pre-Training: MAE. }
At the patch level, we identify smaller anatomical regions or organs. 
 To achieve this, we utilize a masked autoencoder (MAE) that helps the model learn the relationships between patches by reconstructing randomly masked sections of the input images. This process enhances the model's ability to capture intermediate features critical for detailed analysis. Specifically, we initialize the MAE encoder with ViT-Large weights, originally trained on the ImageNet dataset. However, similar to the approach at the image level, we recognize that these weights may lack domain-specific information relevant to medical images. Therefore, before the distillation process, we perform continual pre-training of the MAE encoder on a dataset of 100,000 unlabeled medical images. 

Next, we also use a teacher-student framework with the MAE encoder guiding the SAM encoder with LoRA by minimizing the reconstruction loss.

\noindent \textbf{Pixel-Level Pre-Training: Denoising. }
At the pixel level, the focus is on capturing fine-grained details, such as subtle tissue variations, critical for downstream tasks like segmentation. To achieve this, we utilize a denoising autoencoder, training the model to remove noise from input images. Given the relatively straightforward nature of the denoising task and the lack of large-scale pre-trained weights, we combine the SAM encoder with the U-Net decoder as our denoising model, which is optimized by the reconstruction loss.

\noindent \textbf{Loss Functions of Self-Supervised Pre-Training Stage.}
Despite the differences in detailed implementation across three levels, the underlying principle across all these models remains consistent: the core of each loss function is fundamentally based on reconstruction loss.
\vspace{-1.4em}
\begin{equation}
\mathcal{L}_{\text{recon}}=\frac{1}{n} \sum_{i=1}^n\left\|F\left(x_i\right)-G\left(x_i\right)\right\|^2
\end{equation}
\vspace{-1.6em}

where $n$ is the number of data iteration, $F(*)$ and $G(*)$ represent functions specific to the task. Specifically, at the image and patch levels, $F(*)$  represents the teacher network for input $x_i$, and $G(*)$ is the student network. At pixel level, $ F(*) = \mathds{1}(*)$ indicates that no processing is applied to the input image, $G(*)$ represents the denoising autoencoder.

\begin{table*}[htbp]
\caption{
\textbf{Performance Comparison of SAM and SAM2 Variants on Synapse Dataset. }
\textbf{Bold} numbers indicate the best performance.
By default, we utilize SAM as our base model.
$^{\dagger}$ indicates H-SAM based model; 
$^{\ast}$ indicates SAM2 based model.
\textbf{\textit{ The full table is provided in the Appendix \textcolor{cvprblue}{B.7}.}}
}
\begin{center}
\vspace {-2.0em}
\small
\setlength{\tabcolsep}{3pt}
\begin{tabular}{c|c|c c c c c c c c|c c}
\toprule
\textbf{\makecell[c]{Training\\Set}} & \textbf{Method} & \textbf{Spleen} & \textbf{\makecell[c]{Right\\Kidney}} & \textbf{\makecell[c]{Left\\Kidney} } & \textbf{Gallbladder} & \textbf{Liver} & \textbf{Stomach} & \textbf{Aorta} & \textbf{Pancreas} & \textbf{Mean Dice}  $\textcolor{red}{\uparrow}$ & \textbf{HD} $\textcolor{darkgreen}{\downarrow}$\\
\midrule
\multirow{7}*{10\%} & AutoSAM~\cite{autosam}  & 68.80 & 77.44 & 76.53 & 24.87 & 88.06 & 52.70 & 75.19 & 34.58 & 55.69 & 31.67 \\
~ & SAMed~\cite{samed}  & 85.82 & 82.25 & 82.62 & 63.15 & 92.72 & 67.20 & 78.72 & 52.12 & 75.57 & 23.02 \\
~ & SAMora (Ours) & \textbf{88.04} &\textbf{ 83.41} & \textbf{86.07} & \textbf{67.33} & \textbf{94.27} & \textbf{69.20 }& \textbf{82.85} & \textbf{64.13} & \textbf{79.41} & \textbf{15.68} \\
\cmidrule(lr){2-12} 
~ & SAMed-2$^{\star}$ & 86.61 & 83.01 & 84.56 & 61.51 & 91.07 & 69.02 & 77.99 & 52.09 & 76.68 & 18.93 \\
~ & SAMora-2$^{\star}$ (Ours) & \textbf{87.81} & \textbf{85.73} & \textbf{86.35} & \textbf{68.30} & \textbf{93.78} & \textbf{75.24} & \textbf{81.12} & \textbf{63.62} & \textbf{80.24} & \textbf{16.27}\\
\cmidrule(lr){2-12} 
~ & H-SAM~\cite{hsam} & 90.21 & 84.16 & 85.65 & 70.70 & 94.29 & 76.10 & 85.54 & 56.17 & 80.35 & 15.54 \\
~ & H-SAMora$^{\dagger}$ (Ours) & \textbf{92.46} & \textbf{85.13} & \textbf{86.71} & \textbf{73.15} & \textbf{95.82} & \textbf{81.85} & \textbf{88.56} & \textbf{72.72} & \textbf{84.34} & \textbf{11.63}\\
\midrule        
\multirow{8}*{\makecell[c]{Fully\\Supervised}} 
~ & MERIT~\cite{merit} & 92.01 & 84.85 & 87.79 & 74.40 & 95.26 & 85.38 & 87.71 & 71.81 & 84.90 &  13.22 \\
\cmidrule(lr){2-12} 
~ & SAMed~\cite{samed}  & 87.77 & 69.11 & 80.45 & 79.95 & 94.80 & 72.17 & 88.72 & 82.06 & 81.88 & 20.64 \\
~ & SAMora (Ours) & \textbf{89.27} & \textbf{74.05} & \textbf{81.04} & \textbf{81.51} & \textbf{94.97} & \textbf{74.53} & \textbf{88.87} & \textbf{82.42} & \textbf{83.33} & \textbf{14.57} \\
\cmidrule(lr){2-12} 
~ & SAMed-2$^{\star}$ & 88.63 & 68.63 & 81.22 & 80.33 & 95.18 & 71.00 & 87.63 & 81.93 & 82.12 & 12.76 \\
~ & SAMora-2$^{\star}$ (Ours) & \textbf{91.78} & \textbf{75.85} & \textbf{82.02} & \textbf{83.52} & \textbf{95.49} & \textbf{75.11} & \textbf{87.11} & \textbf{82.26} & \textbf{84.14} & \textbf{10.28}\\
\cmidrule(lr){2-12} 
~ & H-SAM~\cite{hsam} & 93.34 & 89.93 & 91.88 & 73.49 & 95.72 & 87.10 & 89.38 & 71.11 & 86.49 & 8.18 \\
~ & H-SAMora$^{\dagger}$ (Ours) & \textbf{94.62} &\textbf{ 91.45} & \textbf{93.00} & \textbf{76.55} & \textbf{96.51} &\textbf{ 89.95} &\textbf{ 89.55} & \textbf{77.09} & \textbf{88.59} & \textbf{7.09}\\
\bottomrule
\end{tabular}
\label{tab:sam_comparison}
\end{center}
\vspace {-3.0em}
\end{table*}

\subsection{SAMora: Fine-Tuning Stage}
We fine-tune the overview SAM model during this stage, focusing on the decoder, using a smaller labeled dataset.
Furthermore, we design an effective fusion strategy to combine the features from multiple LoRA experts and ensure that each block's strengths are utilized to their fullest potential.

\noindent \textbf{Fusing Multi-LoRA. }
Previous fusion approaches, such as LAC, tend to diminish the unique characteristics of each individual LoRA block, while MoLE does not fully exploit the distinct features present at different hierarchical levels, limiting the model's ability to capture multi-level representations effectively.
Thus, we propose a hierarchical fusion module (HL-Attn) based on a cross-attention mechanism to fuse the features sequentially. 

Referring to Fig.~\ref{fig:hl-attn}, consider a transformer block
of SAM encoder, parameterized by $\theta$ (encompassing both the multi-head attention layer and
the feed-forward neural network), and multiple LoRA experts $\Omega=\{\Delta \theta_{im},\Delta \theta_{pa},\Delta \theta_{pi} \}$, which indicates image-level LoRA, patch-level LoRA, pixel-level LoRA respectively. When given a input $\boldsymbol{x} \in \mathbb{R}^{L \times d}$, the output of the pre-trained model block $\theta$ is presented as $\boldsymbol{F}_\theta \in \mathbb{R}^{L \times d}$ :
\begin{equation}
\boldsymbol{x}_\theta^{\prime}=\boldsymbol{x}+f_{\mathrm{Attn}}(\mathrm{LN}(\boldsymbol{x}) \mid \theta)
\end{equation}
\begin{equation}
\boldsymbol{F}_\theta(\boldsymbol{x})=\boldsymbol{x}_\theta^{\prime}+f_{\mathrm{FFN}}\left(\operatorname{LN}\left(\boldsymbol{x}_\theta^{\prime}\right) \mid \theta\right)
\end{equation}
where $L$ and $d$ indicate the sequence length and the dimension of $\boldsymbol{x}$, respectively. $f_{\mathrm{Attn}}(\cdot)$ and $f_{\mathrm{FFN}}(\cdot)$ denotes the multi-head attention layer and feed-forward neural network, respectively. LN refers to layer normalization. The output of each LoRA is presented as $\boldsymbol{E}_{\Delta \theta_i}(\boldsymbol{x}) \in \mathbb{R}^{L \times d}$.
\begin{equation}
\boldsymbol{x}_{\Delta \theta_i}^{\prime}  =\boldsymbol{x}+f_{\mathrm{Attn}}\left(\mathrm{LN}(\boldsymbol{x}) \mid \Delta \theta_i\right)
\end{equation}
\begin{equation}
\boldsymbol{E}_{\Delta \theta_i}(\boldsymbol{x})  =\boldsymbol{x}_{\Delta \theta_i}^{\prime}+f_{\mathrm{FFN}}\left(\mathrm{LN}\left(\boldsymbol{x}_{\Delta \theta_i}^{\prime}\right) \mid \Delta \theta_i\right)
\end{equation}
After that, we apply HL-Attn to fuse the outputs of multiple LoRA. 
\begin{equation}
\boldsymbol{E}_{\Omega}(\boldsymbol{x}) = f_{\mathrm{HL-Attn}}(\boldsymbol{x}_{\Delta \theta_{im}},\boldsymbol{x}_{\Delta \theta_{pa}},\boldsymbol{x}_{\Delta \theta_{pi}})
\end{equation}
Finally, the final output of this block is computed by adding the output of the HL-Attn to the output of the pre-trained block:
\begin{equation}
\boldsymbol{O}(\boldsymbol{x})=\boldsymbol{F}_\theta(\boldsymbol{x})+\boldsymbol{E}_{\Omega}(\boldsymbol{x}).
\end{equation}

\begin{figure}[t]
    \centering
    \includegraphics[width=1\linewidth]{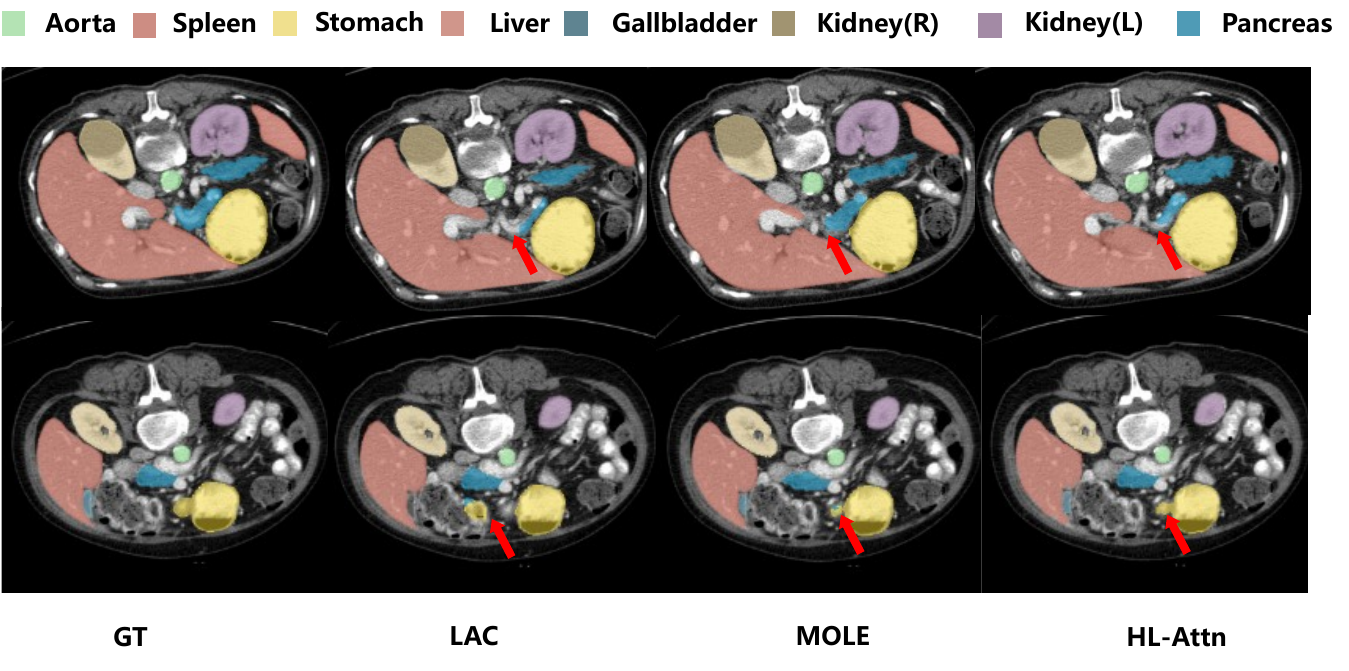}
    \vspace{-2em}
    \caption{\textbf{The Performance of SAMora on Synapse Dataset.}}
    \label{fig:results}
    \vspace{-1.5em}
\end{figure}

\noindent \textbf{Sequential Fusion.}
Given that our approach involves hierarchically fusing LoRA experts from three different levels, it is essential to determine the optimal fusion order. Drawing on insights from DINOv2~\cite{dinov2}, we prioritize the fusion of patch-level and pixel-level features, which capture more fine-grained image information, before incorporating the broader, image-level features.

\noindent \textbf{Cross-Attention.} 
It is particularly well-suited by using cross-attention for this hierarchical fusion strategy because it selectively fuses information across different levels~\cite{crossattn1,crossattn2}. 
Cross-attention mechanisms can facilitate a more effective fusion of diverse representations by dynamically focusing on the most relevant features at each hierarchical level. Specifically, for each fusion step, the feature from the higher-level LoRA block is used as the query (Q), while the features from the lower-level LoRA experts are utilized as the key (K) and value (V). The cross-attention mechanism can then be computed as follows:
\begin{equation}
f_{\mathrm{Cr-Attn}}(Q_H, K_L, V_L)=\operatorname{softmax}\left(\frac{Q_H K^T_L}{\sqrt{d_k}}\right) V_L
\end{equation}
where the $Q_H=W_q \cdot \boldsymbol{E}_{\Delta \theta_H}(\boldsymbol{x}) $,  $K_L=W_k \cdot \boldsymbol{E}_{\Delta \theta_L}(\boldsymbol{x}) $ ,$V_L=W_v \cdot \boldsymbol{E}_{\Delta \theta_L}(\boldsymbol{x}) $, $\boldsymbol{E}_{\Delta \theta_H}(\boldsymbol{x})$ is the feature from higher level LoRA, while the $\boldsymbol{E}_{\Delta \theta_L}(\boldsymbol{x})$ is the feature from lower level LoRA, $d_k$ is the dimension of the key vectors.

\noindent \textbf{Fine-tuning with Labeled Data. } 
Unlike other prompt-free SAM variants, our approach results in distinct LoRA experts after self-supervised stage training, requiring only fine-tuning of the Decoder. 
Specifically, we freeze the SAM encoder and the pre-trained LoRA weights, allowing the HL-Attn module and the subsequent Decoder to be fine-tuned. 
By simplifying the training process in this way, we can focus on optimizing the later stages of the model, reducing overall complexity.

\noindent \noindent \textbf{Flexibility.} 
Although SAMora is primarily built upon the SAMed architecture, it maintains flexibility in adapting to other prompt-free SAM variants. For instance, we combined SAMora with H-SAM and SAMed-2, resulting in two new models, H-SAMora and SAMora-2. Notably, SAMed-2 follows a similar approach to SAMed by applying LoRA to fine-tune the SAM2 model. This adaptability allows SAMora to inherit and integrate improvements from various SAM-based models while ensuring efficient fine-tuning in different medical imaging tasks.

\noindent \textbf{Loss Functions of Fine-Tuning Stage. }
We trained SAMora, H-SAMora and SAMora-2 using the respective loss functions from SAMed, H-SAM and SAMora-2, specifically leveraging dice loss and cross-entropy loss ($L_{dice}$ and $L_{ce}$) as follows, 
\begin{equation}
\mathcal{L}=\lambda_{c e} \mathcal{L}_{c e}+\lambda_{\text {dice }} \mathcal{L}_{\text {dice }}
\end{equation}
where the $L_{dice}$ and $L_{ce}$ denote dice loss and cross-entropy loss, respectively.And the $\lambda_{ce}$ and $\lambda_{dice}$ are set to 0.2 and 0.8, respectively.
\section{Experiment}
\label{sec:experiment}

\subsection{Experimental Setup and Evaluation Metrics}

\noindent \textbf{Datasets. }
For pre-training with unlabeled data, we sampled 100,000 CT images from the Amos22~\cite{amos}, LiTS~\cite{lits}, KiTS~\cite{kits}, and Decathlon Challenge~\cite{deca} datasets. 
The pre-processing steps followed the Fed-MENU~\cite{fed}, and finally, the input images were resized to 224×224 before pre-training.
Furthermore, we conduct fine-tuned experiments on Synapse~\cite{Synapse}, LA~\cite{LA}, and PROMISE12~\cite{promise} datasets. 

\noindent \textbf{Baselines. }
We benchmarked our approach against several prompt-free SAM variants, including MA-SAM~\cite{masam}, I-MedSAM~\cite{imedsam}, AutoSAM~\cite{autosam}, SAM Adapter~\cite{samad}, SAMed~\cite{samed}, and H-SAM~\cite{hsam}. Additionally, we also compared our model with several SOTA methods that are not based on SAM. These include SwinUnet~\cite{Swin-unet}, TransDeepLab~\cite{deeplab}, DAE-Former~\cite{dae}, and MERIT~\cite{merit}. Furthermore, we also compare SAMora and its variants against semi-supervised methods across various datasets, including UA-MT~\cite{uamt}, SS-Net~\cite{ssnet}, MC-Net~\cite{mcnet} and DTC~\cite{dtc}. We followed the data split and preprocessing protocols from H-SAM ~\cite{hsam}. The corresponding part of the dataset was designated as labeled data, while the remaining portion was treated as unlabeled data to facilitate semi-supervised training.

\noindent \textbf{Evaluation Metrics.} 
We utilize the Dice coefficient~\cite{dice} and the average Hausdorff distance (HD)~\cite{hd} as evaluation metrics.
\begin{equation}
\text{Dice} = \frac{2|A \cap B|}{|A| + |B|}, \quad \text{HD} = \frac{1}{N} \sum_{i=1}^{N} d_H(A_i, B_i)
\end{equation}

\noindent \textbf{Evaluation Protocol. }
To ensure a fair comparison, the segmentation result is evaluated on the complete test volumes, following the protocol established by H-SAM~\cite{hsam}.
\begin{itemize}
    \item The Synapse dataset consists of 3,779 contrast-enhanced axial abdominal CT images, with 2,212 slices used in the training set. To effectively demonstrate the efficiency of SAMora, we also fine-tuned the model using only 10\% of the training data. Following the H-SAM protocol, we evaluated the segmentation of eight abdominal organs: aorta, gallbladder, spleen, left kidney, right kidney, liver, pancreas, and stomach. 
    \item The left atrial (LA) dataset is derived from the 2018 Atrial Segmentation Challenge~\cite{LA}. We strictly follow H-SAM for data split and data pre-processing. Specifically, we only keep (4/100)(5\%) scans as labeled data to fine-tune, followed by H-SAM.
    \item PROMISE2012 dataset is derived from the Prostate MR Image Segmentation 2012~\cite{promise}. We strictly follow the data split and pre-processing methods of H-SAM. Specifically, we only keep 7.5\%  (3/40) scans as labeled data to fine-tune followed by H-SAM.
\end{itemize}

\noindent \textbf{Implementation Details. }
All implementations use PyTorch, with all models trained on eight NVIDIA RTX A100 GPUs. The SAM (ViT-B) and SAM2 (hiera-base-plus) backbone are utilized throughout the entire training process. We combine data augmentation techniques, including elastic deformation, rotation, and scaling. The training loss is a combination of Cross-Entropy loss and Dice loss. For all LoRA experts used in this paper, we adopt the same settings as in SAMed and H-SAM in which the rank of LoRA is set to 4. For fairness in comparison, we use the same image resolution of 224×224 on the Synapse dataset, aligning with other SAM variants and SOTA methods. 

\begin{table}[t]
\caption{\textbf{Comparison of SAM Variants against Semi-Supervised Methods across Various Datasets.} \textbf{\textit{The full table is provided in the Appendix \textcolor{cvprblue}{B.7}.}}}
\vspace {-2.0em}
\begin{center}
\small
\setlength{\tabcolsep}{7pt}
\begin{tabular}{c|c|c|c}
\toprule
\textbf{Method} &\textbf{\makecell[c]{10\% \\Synapse}} & \textbf{\makecell[c]{5\%\\ LA}}  & \textbf{\makecell[c]{7.5\% \\PROMISE12}}  \\
\midrule
SS-Net~\cite{ssnet}        & 56.74 & \textbf{86.33}    & \textbf{73.19}      \\
MC-Net~\cite{mcnet}        & \textbf{61.20} & 83.59     & 72.66      \\
\midrule
SAMed~\cite{samed}               & 75.57 & 87.72     & 86.00   \\
SAMora (Ours)               & \textbf{79.41} & \textbf{90.13}  & \textbf{88.44}   \\
\midrule
SAMed-2              & 76.68  & 87.91     & 86.50      \\
SAMora-2 (Ours)             & \textbf{80.24} & \textbf{91.04}  & \textbf{89.27}  \\
\midrule
H-SAM~\cite{hsam}               & 80.35 & 89.22     & 87.27      \\
H-SAMora (Ours)             & \textbf{84.34} & \textbf{92.46}  & \textbf{90.14}  \\
\bottomrule
\end{tabular}
\end{center}
\label{tab:la_pr_results}
\vspace {-3.0em}
\end{table}

\subsection{Results}
Tab.~\ref{tab:sam_comparison} and Tab.~\ref{tab:la_pr_results} compare different SAM variants across various datasets, evaluating their performance using Mean Dice(\%) and HD as the primary metrics. 

\noindent \textbf{Firstly}, SAMora demonstrates exceptional few-shot transferability, achieving impressive results even when fine-tuned on only a fraction of the available data. Compared to other prompt-free SAM variants and other SOTA models, SAMora, SAMora-2, and H-SAMora achieved remarkable Mean Dice scores of 79.41\%, 80.24\%, and 84.34\%, respectively. Furthermore, Tab.~\ref{tab:la_pr_results} shows that SAMora and its variants also achieve the SOTA performance against other SOTA semi-supervised methods across the 10\% Synapse, the 5\% LA and 7.5\% PROMISE12 dataset. 

\noindent \textbf{Secondly}, when evaluated on the full Synapse dataset (100\% Synapse), SAMora and its variants continued to show improvements,
We also conducted \textit{\textbf{statistical validation} }to confirm the significance of our performance improvements in Appendix \textcolor{cvprblue}{B.6}.
 
\noindent \textbf{Thirdly}, Tab.~\ref{tab:fine_tuning} shows the number of fine-tuning epochs and the total parameter among SAMora and its variants. The results indicate that compared to other prompt-free SAM variants, SAMora and H-SAMora demonstrate significantly achieving high performance with considerably 10\% training epochs. 

\noindent \textbf{Visualization.} Fig.~\ref{fig:results} shows the performance of our proposed model, SAMora, on the Synapse dataset. These results highlight the effectiveness of SAMora, further emphasizing its potential as a robust solution for medical image segmentation tasks. \textbf{\textit{Fig.~\textcolor{cvprblue}{6} illustrates the complementary nature of multiple LoRA modules.}}

\begin{table}[t]
\caption{\textbf{Comparison of Fine-tuning Efficiency and Performance of SAMora, SAMora-2, H-SAMora.}}
\vspace {-2.0em}
\begin{center}
\small
\setlength{\tabcolsep}{4pt}
\begin{tabular}{c|c|c|c}
\toprule
\textbf{Method}  & \textbf{\makecell[c]{Fine-tuning \\ Epochs}} & \textbf{\makecell[c]{Total \\ Parameter (M)}}   & \textbf{\makecell[c]{Mean \\Dice (\%)}}  \\
\midrule
SAMed  & 200 & 108.8 & 75.57  \\
SAMora (Ours) & 20 & 118.5 & \textbf{79.41}  \\
\midrule
SAMed-2  & 200 & 99.1 & 76.68  \\
SAMora-2 (Ours) & 20 & 109.7 & \textbf{80.24}  \\
\midrule
 H-SAM  & 300 & 112.3 & 80.35  \\
H-SAMora (Ours)
   & 30 & 122.0 & \textbf{84.34}  \\
\bottomrule
\end{tabular}
\label{tab:fine_tuning}
\end{center}
\vspace {-2.0em}
\end{table}

\begin{table}[t]
\caption{\textbf{Effectiveness of Different Multiple LoRA experts Fusion Strategies.\textit{The full table is provided in the Appendix \textcolor{cvprblue}{B.3}}.}}
\label{tab:effectiveness}
\vspace {-2.0em}
\begin{center}
\small
\setlength{\tabcolsep}{1pt}
\begin{tabular}{c c c c|c}
\toprule
\textbf{\makecell[c]{Image-level\\LoRA}} & \textbf{\makecell[c]{Patch-level\\LoRA}} & \textbf{\makecell[c]{Pixel-level\\LoRA}} & \textbf{\makecell[c]{Fusion\\Module}} & \textbf{\makecell[c]{Mean Dice\\(\%)}} \\
\midrule
\cmark & \cmark & \cmark & LAC~\cite{lac1} & 82.41 \\
\cmark & \cmark & \cmark & MOLE~\cite{mole}& 83.91 \\
\cmark & \cmark & \cmark & LoRAHub~\cite{lorahub} & 81.07 \\
\cmark & \cmark & \cmark & HL-Attn (ours) & \textbf{84.34} \\
\midrule
1 & 1 & 2 & HL-Attn (ours) & 84.21 \\
1 & 2 & 1 & HL-Attn (ours) & 83.86 \\
2 & 1 & 1 & HL-Attn (ours) & \textbf{84.34} \\
\bottomrule
\end{tabular}
\end{center}
\vspace {-3.0em}
\end{table}

\subsection{Model Analysis}


\noindent \textbf{Different LoRA Fusion. } 
As shown in Tab.~\ref{tab:effectiveness} (top) and Fig.~\ref{fig:results}, we compared against two alternative LoRA fusion modules: Linear Arithmetic Composition (LAC)~\cite{lac1, lac2} and Mixture of LoRA Experts (MOLE)~\cite{mole}. 
LAC exhibits the lowest performance, confirming that simple linear weighting is prone to diminishing each trained LoRA's unique characteristics.
In contrast, HL-Attn not only better preserves the individual characteristics of each LoRA but also fuses them hierarchically, leading to superior task performance. \textbf{\textit{The details are further elaborated in Tab.~\textcolor{cvprblue}{10} in Appendix~\textcolor{cvprblue}{B.2}.}}

\noindent \textbf{Sequences of Fusion. }
To illustrate the importance of the fusion sequence in HL-Attn, as shown in Tab.~\ref{tab:effectiveness} (bottom), we conducted experiments evaluating the impact of different LoRA block fusion orders on H-SAMora's performance on the Synapse dataset. 
Here, we compare various fusion sequences, where "1" represents earlier fusion and "2" indicates later fusion of LoRA experts. 
The results show that the sequence where the LoRA experts obey the fusion strategy of 2-1-1 yields the highest Dice of 84.34\%, while the 1-2-1 sequence yields the lowest Dice of 83.86\%. The result demonstrates that the hierarchical fusion strategy indeed influences the performance of models, suggesting that the high-level spatial structures are relatively complex. This underscores the need for an effective approach to fusing various hierarchical features to fully capture the intricacies of these structures.

\begin{figure}[t]
    \centering
    \includegraphics[width=1\linewidth]{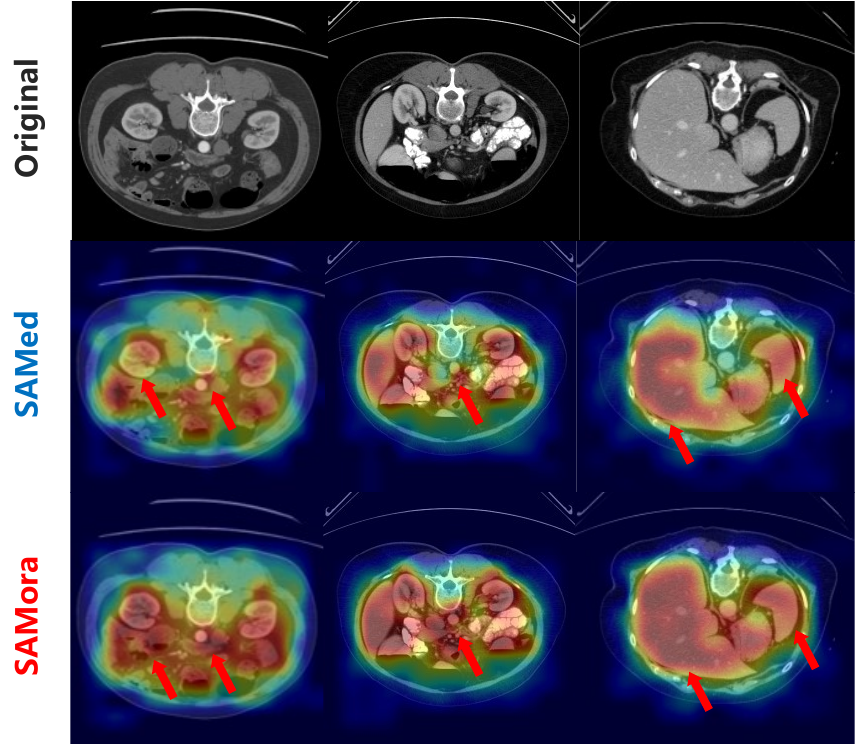}
    \vspace{-2em}
    \caption{\textbf{The Visual Heatmaps between SAMed and SAMora.} The heatmaps display regions of interest with varying levels of relevance, where red denotes areas of high attention, yellow indicates moderate attention, and blue represents low or no attention}
    \label{fig:heatmap}
    \vspace{-1.5em}
\end{figure}

\noindent \textbf{Visual Interpretation.}
To enhance the interpretability of the model and improve transparency in segmentation, Fig.~\ref{fig:heatmap} visualizes the attention maps between SAMed and SAMora. 

On the one hand, the focus regions of SAMed are somewhat scattered, failing not only to fully capture all relevant areas but also covering some irrelevant regions. On the other hand, SAMora exhibits a more focused attention on critical anatomical structures, accurately covering all target organ regions.

\begin{table}[t]
\caption{\textbf{Ablation Analysis of Multiple LoRA experts on 10\% Synapse.} ``Scratch'' means the model is trained from scratch, while ``T-S'' indicates the model is trained by the Teacher-Student framework. \textbf{\textit{The full table is provided in the Appendix \textcolor{cvprblue}{B.7}.}}}
\label{tab:ablation}
\vspace {-2em}
\begin{center}
\small
\setlength{\tabcolsep}{5pt}
\begin{tabular}{c c c|c}
\toprule
\textbf{\makecell[c]{Image-level\\LoRA}} & \textbf{\makecell[c]{Patch-level\\LoRA}}  & \textbf{\makecell[c]{Pixel-level\\LoRA}}  & \textbf{\makecell[c]{Mean Dice\\(\%)}} \\
\midrule
\textbf{Scratch}  & \xmark  & \xmark  & 77.20 \\
\textbf{T-S} ($w/o$ CPT) & \xmark & \xmark  & 77.31 \\
\textbf{T-S} ($w/$ CPT) & \xmark  & \xmark  & \textbf{78.03} \\
\midrule
\xmark  &  \textbf{Scratch} & \xmark  & 76.54 \\
\xmark  &  \textbf{T-S} ($w/o$ CPT) & \xmark & 77.19 \\
\xmark  &  \textbf{T-S} ($w/$ CPT) &  \xmark  & \textbf{78.81} \\
\midrule
\xmark & \xmark &  \textbf{Scratch}   & 76.97 \\
\bottomrule
\end{tabular}
\end{center}
\vspace {-3.25em}
\end{table}

\subsection{Ablation Studies}
\noindent  \textbf{Effectiveness of each LoRA. } In Tab.~\ref{tab:ablation} , we performed ablation studies on 10\% Synapse dataset to evaluate the contribution of each LoRA block at different levels—image, patch, and pixel—to the overall performance of models. By selectively retaining only one LoRA block during fine-tuning, we observed the distinct impact of each level. The results indicate that the patch-level LoRA achieved the highest Mean Dice score of 83.02\%, demonstrating its significant contribution to the model's effectiveness in capturing intermediate-level features. 

\noindent  \textbf{Effectiveness of Teacher-Student Framework. }
We employed distillation techniques based on the Teacher-Student framework to pre-train the SAM encoder at the image and patch level in stage 1. Tab.~\ref{tab:ablation} presents a comparison between the distillation-based training approach and direct training of the SAM encoder. 

In particular, the results indicate that when using the teacher-student framework (T-S) with CPT, the model achieves higher performance (78.81\% and 83.02\%) compared to the models trained directly without distillation. This suggests that distilling knowledge based on Teacher-Student framework can effectively capture more nuanced and hierarchical features within medical images. 

\noindent  \textbf{Effectiveness of Continual Pre-Training. } 
Furthermore, Tab.~\ref{tab:ablation} also highlights the significant impact of CPT during the distillation process. The results clearly demonstrate that models utilizing CPT exhibit substantial improvements compared to those without it, underscoring the importance of this step in enhancing the model's performance. \textbf{\textit{Tab.~\textcolor{cvprblue}{12} in Appendix \textcolor{cvprblue}{B.5} also shows the effectiveness of the CPT.}}

\begin{table}[t]
\label{tab:pcdm}
\vspace{-1.2em}
\begin{center}
  \scriptsize
  \setlength{\tabcolsep}{4pt}
    \caption{\textbf{Performance Comparison of Different Models.}}
    \vspace{-1em}
  \begin{tabular}{l|c c c c c}
    \toprule Model &  MA-SAM & I-MedSAM & SAMora & SAMora-2 & H-SAMora \\
    \midrule
    AMOS22 &  82.70 & 86.26 & 90.51  & 90.77    & \textbf{91.84}     \\
    BTCV   &  83.12 & 85.67& 89.87  & 91.06    &\textbf{92.51}     \\
    Synapse & 72.69 & 75.11 & 79.41 & 80.24 & \textbf{84.34} \\
    \midrule
    Inference Time & 4.3 & 3.4 &\textbf{ 3.1} & 4.2 & 5.7  \\
    \bottomrule
  \end{tabular}
\end{center}
\vspace{-2.4em}
\end{table}

\noindent  \textbf{Comparison of computational efficiency. } 
Tab.\textcolor{cvprblue}{6} presents a performance comparison of different models. Inference time is measured in seconds (s). The results demonstrate that SAMora and its variants achieve an optimal balance between segmentation accuracy and efficiency, consistently outperforming I-MedSAM~\cite{imedsam} and MA-SAM~\cite{masam} across all datasets. Notably, SAMora achieves superior segmentation accuracy while maintaining a lower inference time, reinforcing its practical applicability in real-world scenarios. 

\vspace{-0.5em}
\section{Conclusion}
\label{sec:conclusion}
\vspace{-0.5em}

In this paper, we propose to integrate three hierarchical levels of self-supervised knowledge. Additionally, we designed an HL-Attn fusion module to effectively fuse hierarchical medical knowledge. Our experiments on the LA, PROMISE12, and Synapse datasets in fully supervised and few-shot settings demonstrate that SAMora and its variants consistently outperform other strategies, achieving SOTA performance with a mean Dice score of 84.34\%.

\section*{Acknowledgement}
The authors would like to acknowledge the financial support from
National Science Foundation China grant No. 62173298.

{
    \small
    \bibliographystyle{ieeenat_fullname}
    \bibliography{main}

\begin{thebibliography}{75}
\providecommand{\natexlab}[1]{#1}
\providecommand{\url}[1]{\texttt{#1}}
\expandafter\ifx\csname urlstyle\endcsname\relax
  \providecommand{\doi}[1]{doi: #1}\else
  \providecommand{\doi}{doi: \begingroup \urlstyle{rm}\Url}\fi

\bibitem[Antonelli et~al.(2022)Antonelli, Reinke, Bakas, Farahani, Kopp-Schneider, Landman, Litjens, Menze, Ronneberger, Summers, et~al.]{deca}
Michela Antonelli, Annika Reinke, Spyridon Bakas, Keyvan Farahani, Annette Kopp-Schneider, Bennett~A Landman, Geert Litjens, Bjoern Menze, Olaf Ronneberger, Ronald~M Summers, et~al.
\newblock The medical segmentation decathlon.
\newblock \emph{Nature communications}, 13\penalty0 (1):\penalty0 4128, 2022.

\bibitem[Assran et~al.(2023)Assran, Duval, Misra, Bojanowski, Vincent, Rabbat, LeCun, and Ballas]{I-JEPA}
Mahmoud Assran, Quentin Duval, Ishan Misra, Piotr Bojanowski, Pascal Vincent, Michael Rabbat, Yann LeCun, and Nicolas Ballas.
\newblock Self-supervised learning from images with a joint-embedding predictive architecture.
\newblock In \emph{Proceedings of the IEEE/CVF Conference on Computer Vision and Pattern Recognition}, pages 15619--15629, 2023.

\bibitem[Azad et~al.(2022)Azad, Heidari, Shariatnia, Aghdam, Karimijafarbigloo, Adeli, and Merhof]{deeplab}
Reza Azad, Moein Heidari, Moein Shariatnia, Ehsan~Khodapanah Aghdam, Sanaz Karimijafarbigloo, Ehsan Adeli, and Dorit Merhof.
\newblock Transdeeplab: Convolution-free transformer-based deeplab v3+ for medical image segmentation.
\newblock In \emph{International Workshop on PRedictive Intelligence In MEdicine}, pages 91--102. Springer, 2022.

\bibitem[Azad et~al.(2023)Azad, Arimond, Aghdam, Kazerouni, and Merhof]{dae}
Reza Azad, Ren{\'e} Arimond, Ehsan~Khodapanah Aghdam, Amirhossein Kazerouni, and Dorit Merhof.
\newblock Dae-former: Dual attention-guided efficient transformer for medical image segmentation.
\newblock In \emph{International Workshop on PRedictive Intelligence In MEdicine}, pages 83--95. Springer, 2023.

\bibitem[Bao et~al.(2021)Bao, Dong, Piao, and Wei]{beit}
Hangbo Bao, Li Dong, Songhao Piao, and Furu Wei.
\newblock Beit: Bert pre-training of image transformers.
\newblock \emph{arXiv preprint arXiv:2106.08254}, 2021.

\bibitem[Bilic et~al.(2023)Bilic, Christ, Li, Vorontsov, Ben-Cohen, Kaissis, Szeskin, Jacobs, Mamani, Chartrand, et~al.]{lits}
Patrick Bilic, Patrick Christ, Hongwei~Bran Li, Eugene Vorontsov, Avi Ben-Cohen, Georgios Kaissis, Adi Szeskin, Colin Jacobs, Gabriel Efrain~Humpire Mamani, Gabriel Chartrand, et~al.
\newblock The liver tumor segmentation benchmark (lits).
\newblock \emph{Medical Image Analysis}, 84:\penalty0 102680, 2023.

\bibitem[Cao et~al.(2022)Cao, Wang, Chen, Jiang, Zhang, Tian, and Wang]{Swin-unet}
Hu Cao, Yueyue Wang, Joy Chen, Dongsheng Jiang, Xiaopeng Zhang, Qi Tian, and Manning Wang.
\newblock Swin-unet: Unet-like pure transformer for medical image segmentation.
\newblock In \emph{European conference on computer vision}, pages 205--218. Springer, 2022.

\bibitem[Chen et~al.(2019)Chen, Bai, and Rueckert]{LA}
Chen Chen, Wenjia Bai, and Daniel Rueckert.
\newblock Multi-task learning for left atrial segmentation on ge-mri.
\newblock In \emph{Statistical Atlases and Computational Models of the Heart. Atrial Segmentation and LV Quantification Challenges: 9th International Workshop, STACOM 2018, Held in Conjunction with MICCAI 2018, Granada, Spain, September 16, 2018, Revised Selected Papers 9}, pages 292--301. Springer, 2019.

\bibitem[Chen et~al.(2024{\natexlab{a}})Chen, Miao, Wu, Zhong, Yan, Kim, Hu, Liu, Sun, Li, et~al.]{masam}
Cheng Chen, Juzheng Miao, Dufan Wu, Aoxiao Zhong, Zhiling Yan, Sekeun Kim, Jiang Hu, Zhengliang Liu, Lichao Sun, Xiang Li, et~al.
\newblock Ma-sam: Modality-agnostic sam adaptation for 3d medical image segmentation.
\newblock \emph{Medical Image Analysis}, 98:\penalty0 103310, 2024{\natexlab{a}}.

\bibitem[Chen et~al.(2021)Chen, Lu, Yu, Luo, Adeli, Wang, Lu, Yuille, and Zhou]{transunet}
Jieneng Chen, Yongyi Lu, Qihang Yu, Xiangde Luo, Ehsan Adeli, Yan Wang, Le Lu, Alan~L Yuille, and Yuyin Zhou.
\newblock Transunet: Transformers make strong encoders for medical image segmentation.
\newblock \emph{arXiv preprint arXiv:2102.04306}, 2021.

\bibitem[Chen et~al.(2024{\natexlab{b}})Chen, Wu, Chitta, Jaeger, Geiger, and Li]{autodriving}
Li Chen, Penghao Wu, Kashyap Chitta, Bernhard Jaeger, Andreas Geiger, and Hongyang Li.
\newblock End-to-end autonomous driving: Challenges and frontiers.
\newblock \emph{IEEE Transactions on Pattern Analysis and Machine Intelligence}, 2024{\natexlab{b}}.

\bibitem[Chen et~al.(2020{\natexlab{a}})Chen, Kornblith, Norouzi, and Hinton]{simclr}
Ting Chen, Simon Kornblith, Mohammad Norouzi, and Geoffrey Hinton.
\newblock A simple framework for contrastive learning of visual representations.
\newblock In \emph{International conference on machine learning}, pages 1597--1607. PMLR, 2020{\natexlab{a}}.

\bibitem[Chen et~al.(2020{\natexlab{b}})Chen, Kornblith, Swersky, Norouzi, and Hinton]{simclrv2}
Ting Chen, Simon Kornblith, Kevin Swersky, Mohammad Norouzi, and Geoffrey~E Hinton.
\newblock Big self-supervised models are strong semi-supervised learners.
\newblock \emph{Advances in neural information processing systems}, 33:\penalty0 22243--22255, 2020{\natexlab{b}}.

\bibitem[Chen et~al.(2023{\natexlab{a}})Chen, Zhu, Deng, Cao, Wang, Zhang, Li, Sun, Zang, and Mao]{msa}
Tianrun Chen, Lanyun Zhu, Chaotao Deng, Runlong Cao, Yan Wang, Shangzhan Zhang, Zejian Li, Lingyun Sun, Ying Zang, and Papa Mao.
\newblock Sam-adapter: Adapting segment anything in underperformed scenes.
\newblock In \emph{Proceedings of the IEEE/CVF International Conference on Computer Vision}, pages 3367--3375, 2023{\natexlab{a}}.

\bibitem[Chen et~al.(2023{\natexlab{b}})Chen, Zhu, Ding, Cao, Wang, Li, Sun, Mao, and Zang]{samad}
Tianrun Chen, Lanyun Zhu, Chaotao Ding, Runlong Cao, Yan Wang, Zejian Li, Lingyun Sun, Papa Mao, and Ying Zang.
\newblock Sam fails to segment anything?--sam-adapter: Adapting sam in underperformed scenes: Camouflage, shadow, medical image segmentation, and more.
\newblock \emph{arXiv preprint arXiv:2304.09148}, 2023{\natexlab{b}}.

\bibitem[Cheng et~al.(2024)Cheng, Wei, Zhu, Wang, Qu, Shao, and Zhou]{hsam}
Zhiheng Cheng, Qingyue Wei, Hongru Zhu, Yan Wang, Liangqiong Qu, Wei Shao, and Yuyin Zhou.
\newblock Unleashing the potential of sam for medical adaptation via hierarchical decoding.
\newblock In \emph{Proceedings of the IEEE/CVF Conference on Computer Vision and Pattern Recognition}, pages 3511--3522, 2024.

\bibitem[Cossu et~al.(2024)Cossu, Carta, Passaro, Lomonaco, Tuytelaars, and Bacciu]{cpt1}
Andrea Cossu, Antonio Carta, Lucia Passaro, Vincenzo Lomonaco, Tinne Tuytelaars, and Davide Bacciu.
\newblock Continual pre-training mitigates forgetting in language and vision.
\newblock \emph{Neural Networks}, 179:\penalty0 106492, 2024.

\bibitem[Gao et~al.(2024)Gao, Geng, Zhang, Ma, Fang, Zhang, Li, and Qiao]{adapter}
Peng Gao, Shijie Geng, Renrui Zhang, Teli Ma, Rongyao Fang, Yongfeng Zhang, Hongsheng Li, and Yu Qiao.
\newblock Clip-adapter: Better vision-language models with feature adapters.
\newblock \emph{International Journal of Computer Vision}, 132\penalty0 (2):\penalty0 581--595, 2024.

\bibitem[Gondara(2016)]{denoising1}
Lovedeep Gondara.
\newblock Medical image denoising using convolutional denoising autoencoders.
\newblock In \emph{2016 IEEE 16th international conference on data mining workshops (ICDMW)}, pages 241--246. IEEE, 2016.

\bibitem[Gu et~al.(2024{\natexlab{a}})Gu, Dong, Yang, and Mazurowski]{sam_finetune}
Hanxue Gu, Haoyu Dong, Jichen Yang, and Maciej~A. Mazurowski.
\newblock How to build the best medical image segmentation algorithm using foundation models: a comprehensive empirical study with segment anything model, 2024{\natexlab{a}}.

\bibitem[Gu et~al.(2024{\natexlab{b}})Gu, Wang, Wu, Shi, Chen, Fan, Xiao, Zhao, Chang, Wu, et~al.]{mixofshow}
Yuchao Gu, Xintao Wang, Jay~Zhangjie Wu, Yujun Shi, Yunpeng Chen, Zihan Fan, Wuyou Xiao, Rui Zhao, Shuning Chang, Weijia Wu, et~al.
\newblock Mix-of-show: Decentralized low-rank adaptation for multi-concept customization of diffusion models.
\newblock \emph{Advances in Neural Information Processing Systems}, 36, 2024{\natexlab{b}}.

\bibitem[Hatamizadeh et~al.(2022)Hatamizadeh, Tang, Nath, Yang, Myronenko, Landman, Roth, and Xu]{unetr}
Ali Hatamizadeh, Yucheng Tang, Vishwesh Nath, Dong Yang, Andriy Myronenko, Bennett Landman, Holger~R Roth, and Daguang Xu.
\newblock Unetr: Transformers for 3d medical image segmentation.
\newblock In \emph{Proceedings of the IEEE/CVF winter conference on applications of computer vision}, pages 574--584, 2022.

\bibitem[He et~al.(2016)He, Zhang, Ren, and Sun]{resnet}
Kaiming He, Xiangyu Zhang, Shaoqing Ren, and Jian Sun.
\newblock Deep residual learning for image recognition.
\newblock In \emph{Proceedings of the IEEE conference on computer vision and pattern recognition}, pages 770--778, 2016.

\bibitem[He et~al.(2020)He, Fan, Wu, Xie, and Girshick]{moco}
Kaiming He, Haoqi Fan, Yuxin Wu, Saining Xie, and Ross Girshick.
\newblock Momentum contrast for unsupervised visual representation learning.
\newblock In \emph{Proceedings of the IEEE/CVF conference on computer vision and pattern recognition}, pages 9729--9738, 2020.

\bibitem[He et~al.(2022)He, Chen, Xie, Li, Doll{\'a}r, and Girshick]{mae}
Kaiming He, Xinlei Chen, Saining Xie, Yanghao Li, Piotr Doll{\'a}r, and Ross Girshick.
\newblock Masked autoencoders are scalable vision learners.
\newblock In \emph{Proceedings of the IEEE/CVF conference on computer vision and pattern recognition}, pages 16000--16009, 2022.

\bibitem[Heller et~al.(2019)Heller, Sathianathen, Kalapara, Walczak, Moore, Kaluzniak, Rosenberg, Blake, Rengel, Oestreich, et~al.]{kits}
Nicholas Heller, Niranjan Sathianathen, Arveen Kalapara, Edward Walczak, Keenan Moore, Heather Kaluzniak, Joel Rosenberg, Paul Blake, Zachary Rengel, Makinna Oestreich, et~al.
\newblock The kits19 challenge data: 300 kidney tumor cases with clinical context, ct semantic segmentations, and surgical outcomes.
\newblock \emph{arXiv preprint arXiv:1904.00445}, 2019.

\bibitem[Hu et~al.(2022)Hu, Li, Liu, Chen, Wang, and Liu]{ts}
Chengming Hu, Xuan Li, Dan Liu, Xi Chen, Ju Wang, and Xue Liu.
\newblock Teacher-student architecture for knowledge learning: A survey.
\newblock \emph{arXiv preprint arXiv:2210.17332}, 2022.

\bibitem[Hu et~al.(2021)Hu, Shen, Wallis, Allen-Zhu, Li, Wang, Wang, and Chen]{lora}
Edward~J Hu, Yelong Shen, Phillip Wallis, Zeyuan Allen-Zhu, Yuanzhi Li, Shean Wang, Lu Wang, and Weizhu Chen.
\newblock Lora: Low-rank adaptation of large language models.
\newblock \emph{arXiv preprint arXiv:2106.09685}, 2021.

\bibitem[Huang et~al.(2023{\natexlab{a}})Huang, Liu, Lin, Pang, Du, and Lin]{lac2}
Chengsong Huang, Qian Liu, Bill~Yuchen Lin, Tianyu Pang, Chao Du, and Min Lin.
\newblock Lorahub: Efficient cross-task generalization via dynamic lora composition.
\newblock \emph{arXiv preprint arXiv:2307.13269}, 2023{\natexlab{a}}.

\bibitem[Huang et~al.(2023{\natexlab{b}})Huang, Liu, Lin, Pang, Du, and Lin]{lorahub}
Chengsong Huang, Qian Liu, Bill~Yuchen Lin, Tianyu Pang, Chao Du, and Min Lin.
\newblock Lorahub: Efficient cross-task generalization via dynamic lora composition.
\newblock \emph{arXiv preprint arXiv:2307.13269}, 2023{\natexlab{b}}.

\bibitem[Huang et~al.(2024)Huang, Yang, Liu, Zhou, Chang, Zhou, Chen, Yu, Chen, Chen, et~al.]{huang2024segment}
Yuhao Huang, Xin Yang, Lian Liu, Han Zhou, Ao Chang, Xinrui Zhou, Rusi Chen, Junxuan Yu, Jiongquan Chen, Chaoyu Chen, et~al.
\newblock Segment anything model for medical images?
\newblock \emph{Medical Image Analysis}, 92:\penalty0 103061, 2024.

\bibitem[Huo et~al.(2024)Huo, Sun, Tian, Wang, Yu, Long, Zhang, and Li]{hifuse}
Xiangzuo Huo, Gang Sun, Shengwei Tian, Yan Wang, Long Yu, Jun Long, Wendong Zhang, and Aolun Li.
\newblock Hifuse: Hierarchical multi-scale feature fusion network for medical image classification.
\newblock \emph{Biomedical Signal Processing and Control}, 87:\penalty0 105534, 2024.

\bibitem[Isensee et~al.(2021)Isensee, Jaeger, Kohl, Petersen, and Maier-Hein]{nnUnet}
Fabian Isensee, Paul~F Jaeger, Simon~AA Kohl, Jens Petersen, and Klaus~H Maier-Hein.
\newblock nnu-net: a self-configuring method for deep learning-based biomedical image segmentation.
\newblock \emph{Nature methods}, 18\penalty0 (2):\penalty0 203--211, 2021.

\bibitem[Ji et~al.(2022)Ji, Bai, Ge, Yang, Zhu, Zhang, Li, Zhanng, Ma, Wan, et~al.]{amos}
Yuanfeng Ji, Haotian Bai, Chongjian Ge, Jie Yang, Ye Zhu, Ruimao Zhang, Zhen Li, Lingyan Zhanng, Wanling Ma, Xiang Wan, et~al.
\newblock Amos: A large-scale abdominal multi-organ benchmark for versatile medical image segmentation.
\newblock \emph{Advances in neural information processing systems}, 35:\penalty0 36722--36732, 2022.

\bibitem[Kirillov et~al.(2023)Kirillov, Mintun, Ravi, Mao, Rolland, Gustafson, Xiao, Whitehead, Berg, Lo, et~al.]{sam}
Alexander Kirillov, Eric Mintun, Nikhila Ravi, Hanzi Mao, Chloe Rolland, Laura Gustafson, Tete Xiao, Spencer Whitehead, Alexander~C Berg, Wan-Yen Lo, et~al.
\newblock Segment anything.
\newblock In \emph{Proceedings of the IEEE/CVF International Conference on Computer Vision}, pages 4015--4026, 2023.

\bibitem[Landman et~al.(2015)Landman, Xu, Igelsias, Styner, Langerak, and Klein]{Synapse}
Bennett Landman, Zhoubing Xu, J Igelsias, Martin Styner, Thomas Langerak, and Arno Klein.
\newblock Miccai multi-atlas labeling beyond the cranial vault--workshop and challenge.
\newblock In \emph{Proc. MICCAI Multi-Atlas Labeling Beyond Cranial Vault—Workshop Challenge}, page~12, 2015.

\bibitem[Li et~al.(2023)Li, Khanduri, Qiang, Ibn~Sultan, Chetty, and Zhu]{autosam}
Chengyin Li, Prashant Khanduri, Yao Qiang, Rafi Ibn~Sultan, Indrin Chetty, and Dongxiao Zhu.
\newblock Auto-prompting sam for mobile friendly 3d medical image segmentation.
\newblock \emph{arXiv e-prints}, pages arXiv--2308, 2023.

\bibitem[Lin et~al.(2024)Lin, Chen, Feng, and Huang]{hierarchical1}
Cong Lin, Yinjie Chen, Siling Feng, and Mengxing Huang.
\newblock A multibranch and multiscale neural network based on semantic perception for multimodal medical image fusion.
\newblock \emph{Scientific Reports}, 14\penalty0 (1):\penalty0 17609, 2024.

\bibitem[Lin et~al.(2022)Lin, Cheng, Wu, and Shen]{crossattn1}
Hezheng Lin, Xing Cheng, Xiangyu Wu, and Dong Shen.
\newblock Cat: Cross attention in vision transformer.
\newblock In \emph{2022 IEEE international conference on multimedia and expo (ICME)}, pages 1--6. IEEE, 2022.

\bibitem[Litjens et~al.(2014)Litjens, Toth, Van De~Ven, Hoeks, Kerkstra, Van~Ginneken, Vincent, Guillard, Birbeck, Zhang, et~al.]{promise}
Geert Litjens, Robert Toth, Wendy Van De~Ven, Caroline Hoeks, Sjoerd Kerkstra, Bram Van~Ginneken, Graham Vincent, Gwenael Guillard, Neil Birbeck, Jindang Zhang, et~al.
\newblock Evaluation of prostate segmentation algorithms for mri: the promise12 challenge.
\newblock \emph{Medical image analysis}, 18\penalty0 (2):\penalty0 359--373, 2014.

\bibitem[Liu et~al.(2024)Liu, Xu, Gao, Li, Wang, Chabin, Oguz, and Grbic]{part2}
Han Liu, Zhoubing Xu, Riqiang Gao, Hao Li, Jianing Wang, Guillaume Chabin, Ipek Oguz, and Sasa Grbic.
\newblock Cosst: Multi-organ segmentation with partially labeled datasets using comprehensive supervisions and self-training.
\newblock \emph{IEEE Transactions on Medical Imaging}, 2024.

\bibitem[Luo et~al.(2021)Luo, Chen, Song, and Wang]{dtc}
Xiangde Luo, Jieneng Chen, Tao Song, and Guotai Wang.
\newblock Semi-supervised medical image segmentation through dual-task consistency.
\newblock In \emph{Proceedings of the AAAI conference on artificial intelligence}, pages 8801--8809, 2021.

\bibitem[Ma et~al.(2021)Ma, Zhang, Gu, Zhu, Ge, Zhang, An, Wang, Wang, Liu, et~al.]{ma2024segment13}
Jun Ma, Yao Zhang, Song Gu, Cheng Zhu, Cheng Ge, Yichi Zhang, Xingle An, Congcong Wang, Qiyuan Wang, Xin Liu, et~al.
\newblock Abdomenct-1k: Is abdominal organ segmentation a solved problem?
\newblock \emph{IEEE Transactions on Pattern Analysis and Machine Intelligence}, 44\penalty0 (10):\penalty0 6695--6714, 2021.

\bibitem[Mazurowski et~al.(2023)Mazurowski, Dong, Gu, Yang, Konz, and Zhang]{sam-ft}
Maciej~A Mazurowski, Haoyu Dong, Hanxue Gu, Jichen Yang, Nicholas Konz, and Yixin Zhang.
\newblock Segment anything model for medical image analysis: an experimental study.
\newblock \emph{Medical Image Analysis}, 89:\penalty0 102918, 2023.

\bibitem[Oquab et~al.(2023)Oquab, Darcet, Moutakanni, Vo, Szafraniec, Khalidov, Fernandez, Haziza, Massa, El-Nouby, et~al.]{dinov2}
Maxime Oquab, Timoth{\'e}e Darcet, Th{\'e}o Moutakanni, Huy Vo, Marc Szafraniec, Vasil Khalidov, Pierre Fernandez, Daniel Haziza, Francisco Massa, Alaaeldin El-Nouby, et~al.
\newblock Dinov2: Learning robust visual features without supervision.
\newblock \emph{arXiv preprint arXiv:2304.07193}, 2023.

\bibitem[Ostapenko et~al.(2024)Ostapenko, Su, Ponti, Charlin, Roux, Pereira, Caccia, and Sordoni]{arrow}
Oleksiy Ostapenko, Zhan Su, Edoardo~Maria Ponti, Laurent Charlin, Nicolas~Le Roux, Matheus Pereira, Lucas Caccia, and Alessandro Sordoni.
\newblock Towards modular llms by building and reusing a library of loras.
\newblock \emph{arXiv preprint arXiv:2405.11157}, 2024.

\bibitem[Petit et~al.(2021)Petit, Thome, and Soler]{part1}
Olivier Petit, Nicolas Thome, and Luc Soler.
\newblock Iterative confidence relabeling with deep convnets for organ segmentation with partial labels.
\newblock \emph{Computerized Medical Imaging and Graphics}, 91:\penalty0 101938, 2021.

\bibitem[Rahman and Marculescu(2024)]{merit}
Md~Mostafijur Rahman and Radu Marculescu.
\newblock Multi-scale hierarchical vision transformer with cascaded attention decoding for medical image segmentation.
\newblock In \emph{Medical Imaging with Deep Learning}, pages 1526--1544. PMLR, 2024.

\bibitem[Ravi et~al.(2024)Ravi, Gabeur, Hu, Hu, Ryali, Ma, Khedr, R{\"a}dle, Rolland, Gustafson, et~al.]{sam2}
Nikhila Ravi, Valentin Gabeur, Yuan-Ting Hu, Ronghang Hu, Chaitanya Ryali, Tengyu Ma, Haitham Khedr, Roman R{\"a}dle, Chloe Rolland, Laura Gustafson, et~al.
\newblock Sam 2: Segment anything in images and videos.
\newblock \emph{arXiv preprint arXiv:2408.00714}, 2024.

\bibitem[Shaker et~al.(2024)Shaker, Maaz, Rasheed, Khan, Yang, and Khan]{unetr++}
Abdelrahman~M Shaker, Muhammad Maaz, Hanoona Rasheed, Salman Khan, Ming-Hsuan Yang, and Fahad~Shahbaz Khan.
\newblock Unetr++: delving into efficient and accurate 3d medical image segmentation.
\newblock \emph{IEEE Transactions on Medical Imaging}, 2024.

\bibitem[Shamir et~al.(2019)Shamir, Duchin, Kim, Sapiro, and Harel]{dice}
Reuben~R Shamir, Yuval Duchin, Jinyoung Kim, Guillermo Sapiro, and Noam Harel.
\newblock Continuous dice coefficient: a method for evaluating probabilistic segmentations.
\newblock \emph{arXiv preprint arXiv:1906.11031}, 2019.

\bibitem[Sun et~al.(2020)Sun, Wang, Li, Feng, Tian, Wu, and Wang]{cpt2}
Yu Sun, Shuohuan Wang, Yukun Li, Shikun Feng, Hao Tian, Hua Wu, and Haifeng Wang.
\newblock Ernie 2.0: A continual pre-training framework for language understanding.
\newblock In \emph{Proceedings of the AAAI conference on artificial intelligence}, pages 8968--8975, 2020.

\bibitem[Taha and Hanbury(2015)]{hd}
Abdel~Aziz Taha and Allan Hanbury.
\newblock An efficient algorithm for calculating the exact hausdorff distance.
\newblock \emph{IEEE transactions on pattern analysis and machine intelligence}, 37\penalty0 (11):\penalty0 2153--2163, 2015.

\bibitem[Tian et~al.(2020)Tian, Sun, Poole, Krishnan, Schmid, and Isola]{cl}
Yonglong Tian, Chen Sun, Ben Poole, Dilip Krishnan, Cordelia Schmid, and Phillip Isola.
\newblock What makes for good views for contrastive learning?
\newblock \emph{Advances in neural information processing systems}, 33:\penalty0 6827--6839, 2020.

\bibitem[Trujillano et~al.(2024)Trujillano, Jimenez, Manrique, Kahamba, Okumu, Apollinaire, Carrasco-Escobar, Barrett, and Fornace]{sate}
Fedra Trujillano, Gabriel Jimenez, Edgar Manrique, Najat~F Kahamba, Fredros Okumu, Nombre Apollinaire, Gabriel Carrasco-Escobar, Brian Barrett, and Kimberly Fornace.
\newblock Using image segmentation models to analyse high-resolution earth observation data: new tools to monitor disease risks in changing environments.
\newblock \emph{International Journal of Health Geographics}, 23\penalty0 (1):\penalty0 13, 2024.

\bibitem[Vaswani(2017)]{crossattn2}
A Vaswani.
\newblock Attention is all you need.
\newblock \emph{Advances in Neural Information Processing Systems}, 2017.

\bibitem[Wang et~al.(2024)Wang, Ping, Wang, Han, Chen, Liu, and Sun]{loraflow}
Hanqing Wang, Bowen Ping, Shuo Wang, Xu Han, Yun Chen, Zhiyuan Liu, and Maosong Sun.
\newblock Lora-flow: Dynamic lora fusion for large language models in generative tasks.
\newblock \emph{arXiv preprint arXiv:2402.11455}, 2024.

\bibitem[Wei et~al.(2023)Wei, Cao, Jin, Lu, Wang, and Zhang]{wei2023medsam}
Xiaobao Wei, Jiajun Cao, Yizhu Jin, Ming Lu, Guangyu Wang, and Shanghang Zhang.
\newblock I-medsam: Implicit medical image segmentation with segment anything.
\newblock \emph{arXiv preprint arXiv:2311.17081}, 2023.

\bibitem[Wei et~al.(2024)Wei, Cao, Jin, Lu, Wang, and Zhang]{imedsam}
Xiaobao Wei, Jiajun Cao, Yizhu Jin, Ming Lu, Guangyu Wang, and Shanghang Zhang.
\newblock I-medsam: Implicit medical image segmentation with segment anything.
\newblock In \emph{European Conference on Computer Vision}, pages 90--107. Springer, 2024.

\bibitem[Wen et~al.(2024)Wen, Yuan, Ni, Xiao, and Wu]{denoising2}
Ruxue Wen, Hangjie Yuan, Dong Ni, Wenbo Xiao, and Yaoyao Wu.
\newblock From denoising training to test-time adaptation: Enhancing domain generalization for medical image segmentation.
\newblock In \emph{Proceedings of the IEEE/CVF Winter Conference on Applications of Computer Vision}, pages 464--474, 2024.

\bibitem[Wu et~al.(2021{\natexlab{a}})Wu, Wu, and Huang]{infonce}
Chuhan Wu, Fangzhao Wu, and Yongfeng Huang.
\newblock Rethinking infonce: How many negative samples do you need?
\newblock \emph{arXiv preprint arXiv:2105.13003}, 2021{\natexlab{a}}.

\bibitem[Wu et~al.(2024)Wu, Huang, and Wei]{mole}
Xun Wu, Shaohan Huang, and Furu Wei.
\newblock Mixture of lora experts.
\newblock \emph{arXiv preprint arXiv:2404.13628}, 2024.

\bibitem[Wu et~al.(2021{\natexlab{b}})Wu, Xu, Ge, Cai, and Zhang]{mcnet}
Yicheng Wu, Minfeng Xu, Zongyuan Ge, Jianfei Cai, and Lei Zhang.
\newblock Semi-supervised left atrium segmentation with mutual consistency training.
\newblock In \emph{Medical image computing and computer assisted intervention--MICCAI 2021: 24th international conference, Strasbourg, France, September 27--October 1, 2021, proceedings, part II 24}, pages 297--306. Springer, 2021{\natexlab{b}}.

\bibitem[Wu et~al.(2022)Wu, Wu, Wu, Ge, and Cai]{ssnet}
Yicheng Wu, Zhonghua Wu, Qianyi Wu, Zongyuan Ge, and Jianfei Cai.
\newblock Exploring smoothness and class-separation for semi-supervised medical image segmentation.
\newblock In \emph{International conference on medical image computing and computer-assisted intervention}, pages 34--43. Springer, 2022.

\bibitem[Xiong et~al.(2024)Xiong, Varadarajan, Wu, Xiang, Xiao, Zhu, Dai, Wang, Sun, Iandola, et~al.]{efficientsam}
Yunyang Xiong, Bala Varadarajan, Lemeng Wu, Xiaoyu Xiang, Fanyi Xiao, Chenchen Zhu, Xiaoliang Dai, Dilin Wang, Fei Sun, Forrest Iandola, et~al.
\newblock Efficientsam: Leveraged masked image pretraining for efficient segment anything.
\newblock In \emph{Proceedings of the IEEE/CVF Conference on Computer Vision and Pattern Recognition}, pages 16111--16121, 2024.

\bibitem[Xu et~al.(2023)Xu, Deng, Gateno, and Yan]{fed}
Xuanang Xu, Hannah~H Deng, Jamie Gateno, and Pingkun Yan.
\newblock Federated multi-organ segmentation with inconsistent labels.
\newblock \emph{IEEE transactions on medical imaging}, 42\penalty0 (10):\penalty0 2948--2960, 2023.

\bibitem[Xu et~al.(2024)Xu, Zhu, and Yang]{xu2024mc}
Yunqiu Xu, Linchao Zhu, and Yi Yang.
\newblock Mc-bench: A benchmark for multi-context visual grounding in the era of mllms.
\newblock \emph{arXiv preprint arXiv:2410.12332}, 2024.

\bibitem[Yang et~al.(2024)Yang, Bi, Zhang, and Sun]{samunet}
Sihan Yang, Haixia Bi, Hai Zhang, and Jian Sun.
\newblock Sam-unet: Enhancing zero-shot segmentation of sam for universal medical images.
\newblock \emph{arXiv preprint arXiv:2408.09886}, 2024.

\bibitem[Yang et~al.(2023)Yang, Wu, He, Zhao, and Liu]{sam3d}
Yunhan Yang, Xiaoyang Wu, Tong He, Hengshuang Zhao, and Xihui Liu.
\newblock Sam3d: Segment anything in 3d scenes.
\newblock \emph{arXiv preprint arXiv:2306.03908}, 2023.

\bibitem[Yu et~al.(2019)Yu, Wang, Li, Fu, and Heng]{uamt}
Lequan Yu, Shujun Wang, Xiaomeng Li, Chi-Wing Fu, and Pheng-Ann Heng.
\newblock Uncertainty-aware self-ensembling model for semi-supervised 3d left atrium segmentation.
\newblock In \emph{Medical image computing and computer assisted intervention--MICCAI 2019: 22nd international conference, Shenzhen, China, October 13--17, 2019, proceedings, part II 22}, pages 605--613. Springer, 2019.

\bibitem[Zhang et~al.(2023{\natexlab{a}})Zhang, Liu, He, et~al.]{lac1}
Jinghan Zhang, Junteng Liu, Junxian He, et~al.
\newblock Composing parameter-efficient modules with arithmetic operation.
\newblock \emph{Advances in Neural Information Processing Systems}, 36:\penalty0 12589--12610, 2023{\natexlab{a}}.

\bibitem[Zhang et~al.(2023{\natexlab{b}})Zhang, Ma, Kapse, Saltz, Vakalopoulou, Prasanna, and Samaras]{sampath}
Jingwei Zhang, Ke Ma, Saarthak Kapse, Joel Saltz, Maria Vakalopoulou, Prateek Prasanna, and Dimitris Samaras.
\newblock Sam-path: A segment anything model for semantic segmentation in digital pathology.
\newblock In \emph{International Conference on Medical Image Computing and Computer-Assisted Intervention}, pages 161--170. Springer, 2023{\natexlab{b}}.

\bibitem[Zhang and Liu(2023)]{samed}
Kaidong Zhang and Dong Liu.
\newblock Customized segment anything model for medical image segmentation.
\newblock \emph{arXiv preprint arXiv:2304.13785}, 2023.

\bibitem[Zhao et~al.(2024)Zhao, Gan, Wang, Zhou, Yang, Kuang, and Wu]{loraretriever}
Ziyu Zhao, Leilei Gan, Guoyin Wang, Wangchunshu Zhou, Hongxia Yang, Kun Kuang, and Fei Wu.
\newblock Loraretriever: Input-aware lora retrieval and composition for mixed tasks in the wild.
\newblock \emph{arXiv preprint arXiv:2402.09997}, 2024.

\bibitem[Zhou et~al.(2021)Zhou, Guo, Zhang, Yu, Wang, and Yu]{nnformer}
Hong-Yu Zhou, Jiansen Guo, Yinghao Zhang, Lequan Yu, Liansheng Wang, and Yizhou Yu.
\newblock nnformer: Interleaved transformer for volumetric segmentation.
\newblock \emph{arXiv preprint arXiv:2109.03201}, 2021.

\end{thebibliography}
}

\clearpage
\setcounter{page}{1}
\maketitlesupplementary

\appendix
\setcounter{figure}{5}
\setcounter{table}{6} 
In this supplementary material, we first provide more implementation details for
training strategies and datasets (Sec.~\ref{sec:implement}). Then, we conduct more
additional ablation studies (Sec.~\ref{sec:ablation}) to validate the
effectiveness of each component in our proposed method. Finally, we discuss
SAMora's limitations and potential directions for future work (Sec.~\ref{supsec:limit}).

\section{Implementation Details}
\label{sec:implement}
\subsection{Training strategy}
We provide the training strategy and hyperparameter settings as supplementary material.

In Stage 1, we perform pretraining for image-level, patch-level, and pixel-level
tasks using different models: SimCLRv2 (ResNet50 (2X+SK)) for the image-level task,
MAE (ViT-Large) for the patch-level task, and a denoising model (U-net model) for
the pixel-level task. As shown in Table 6, for image-level task, We adopt warmup
during training, The learning rate is linearly increased for the first 5\% of
epochs, and then decayed with a cosine decay schedule where the weight decay is $1
e^{-4}$, followed by SimCLRv2~\cite{simclrv2}. For the patch-level and pixel-level
tasks, we use the AdamW optimizer. The optimizer momentum is set to 0.9 and 0.95
for the patch-level task, and 0.9 and 0.99 for the pixel-level task, respectively.

\begin{table}[t]
   \caption{Stage1 Setting.}
   \vspace{-2.0em}
   \begin{center}
      \small
      \setlength{\tabcolsep}{6pt}
      \begin{tabular}{c|c|c|c}
         \toprule \textbf{Config} & \textbf{image-level} & \textbf{patch-level} & \textbf{pixel-level} \\
         \midrule Optimizer       & LARS                 & AdamW                & AdamW                \\
         Base learning rate       & 0.075                & 1.5e-4               & 1e-4                 \\
         Batch size               & 512                  & 512                  & 512                  \\
         Weight decay             & 1e-4                 & 0.05                 & 0.05                 \\
         Warmup period            & 30                   & 30                   & 10                   \\
         epoch nums               & 80                   & 60                   & 30                   \\
         \bottomrule
      \end{tabular}
      \label{tab:rank}
   \end{center}
   \vspace{-2.0em}
\end{table}

The training loss is a combination of Dice loss and Mean Squared Error (MSE)
loss. As indicated in Tab.~\textcolor{cvprblue}{7}, the weights for these losses are set to 0.9 for
Dice loss and 0.1 for MSE loss. In our two-stage hierarchical structure, each
stage applies a weighted loss, controlled by a parameter that gradually decreases
through exponential decay, starting from 0.4 and reaching 0 over 300 epochs.

In Tab.~\textcolor{cvprblue}{8}, we present the settings for Stage 2 across various backbones. For
SAMora, SAMed (ViT-B) serves as the backbone. The loss weights are assigned as
0.2 for cross-entropy and 0.8 for Dice loss. For the warmup configuration, the
initial learning rate is set to 0.005, with a warmup period of 250 steps, and
the total number of iterations is 18,600. Notably,
the learning rate adjustment strategy is described as follows:

\begin{equation}
   l r=\left\{
   \begin{array}{lr}
      T \frac{I_{l r}}{WP},                 & T<=WP, \\
      I_{lr}\left(1-\frac{T-WP}{MI}\right), & T>WP .
   \end{array}\right.
\end{equation}

Where $I_{lr}$ represents the initial learning rate, while $T$, $WP$, and $MI$
denote the training iterations, warmup period, and maximum iterations, respectively.

SAMora-2 uses SAM2 (hiera-base-plus) as the backbone, with the only difference
being that the number of epochs is set to 25. All other training parameters
remain the same as those of SAMora. The configuration of H-SAMora follows the guidelines
of H-SAM~\cite{hsam}.

\begin{table}[t]
   \caption{Stage2 Setting.}
   \vspace{-2.0em}
   \begin{center}
      \small
      \setlength{\tabcolsep}{6pt}
      \begin{tabular}{c|c|c|c}
         \toprule \textbf{Config} & \textbf{SAMora} & \textbf{SAMora-2} & \textbf{H-SAMora} \\
         \midrule Optimizer       & AdamW           & AdamW             & AdamW             \\
         Base learning rate       & 5e-3            & 5e-3              & 2.5e-3            \\
         Batch size               & 32              & 32                & 32                \\
         Weight decay             & 0.1             & 0.1               & 0.1               \\
         Warmup period            & 25              & 25                & 25                \\
         epoch nums               & 20              & 25                & 30                \\
         \bottomrule
      \end{tabular}
      \label{tab:rank}
   \end{center}
   \vspace{-2.0em}
\end{table}

\subsection{Additional datasets information}
We detail the dataset settings. Firstly, the unlabeled data that we use to pre-train
is sampled from the Amos22~\cite{amos}, LiTS~\cite{lits}, KiTS~\cite{kits}, and Decathlon
Challenge~\cite{deca} datasets.

\begin{itemize}
   \item AMOS22~\cite{amos} is a large-scale dataset that provides 500 CT and 100
      MRI scans with voxel-level annotations for 15 abdominal organs, supporting
      both CT-only and cross-modality segmentation tasks across diverse clinical
      scenarios.

   \item The LiTS dataset ~\cite{lits} focuses on liver and liver tumor
      segmentation. It comprises 201 abdominal CT volumes, helping to tackle
      challenges such as lesion variability and segmentation complexity, making it
      a widely used benchmark for medical imaging algorithms.

   \item The KiTS dataset~\cite{kits} emphasizes kidney and kidney tumor segmentation.
      Its 2019 release, KiTS19, includes 300 CT cases, collected from patients who
      underwent nephrectomy, and is designed to support automated kidney and
      tumor segmentation research through comprehensive annotations.

   \item The Decathlon Challenge dataset~\cite{deca} offers a broad range of segmentation
      tasks across multiple organs, aiming to advance generalization in medical
      image analysis. It provides an opportunity to test algorithms on various anatomical
      regions and imaging scenarios, making it ideal for benchmarking
      segmentation models across different tasks.
\end{itemize}

In Stage 2, we utilize the Synapse dataset from the MICCAI 2015 Multi-Atlas Abdomen
Labeling Challenge. For the fully supervised training setup, we adhere to the H-SAM
framework to evaluate the segmentation performance across eight abdominal organs:
the aorta, gallbladder, spleen, left kidney, right kidney, liver, pancreas, and stomach.

In addition to the fully supervised setup, we also implement a few-shot learning
scenario. For this, we adopt a slice-based data selection strategy, randomly sampling
10\% of the training data (221 slices) from different subjects within the complete training set, which consists of 2,212 axial slices.

\subsection{Preprocessing and augmentation strategies for training datasets}
\begin{table}[t]
   \caption{Ablation study on rank size of LoRA layers.}
   \vspace{-2.0em}
   \begin{center}
      \small
      \setlength{\tabcolsep}{8pt}
      \begin{tabular}{c|c|c|c}
         \toprule \textbf{Method} & \textbf{Rank = 1} & \textbf{Rank = 4} & \textbf{Rank = 16} \\
         \midrule SAMed           & 69.12             & 75.57             & 69.03              \\
         SAMora (Ours)            & 75.26             & 79.41             & 76.88              \\
         \midrule SAMed-2         & 69.89             & 76.68             & 73.54              \\
         SAMora-2 (Ours)          & 75.53             & 80.24             & 76.12              \\
         \midrule H-SAM           & 72.14             & 80.35             & 77.14              \\
         H-SAMora (Ours)           
                                  & \textbf{78.91}    & \textbf{84.34}    & \textbf{80.57}     \\
         \bottomrule
      \end{tabular}
      \label{tab:rank}
   \end{center}
   \vspace{-2.0em}
\end{table}

To improve the generalization ability of the model and enhance the robustness of training, we follow the preprocessing and data augmentation strategies adopted in TransUNet~\cite{transunet}, SAMed~\cite{samed}, H-SAM~\cite{hsam}. 

The original medical images are first resampled to a uniform spatial resolution to mitigate variations caused by different imaging protocols. Following TransUNet, for 3D volumetric data, each volume is processed in a slice-by-slice manner, where the slices are extracted along the axial plane. The extracted 2D slices are then normalized to zero mean and unit variance to ensure consistent intensity distributions across different datasets.

To prevent overfitting and improve the diversity of training samples, we employ several data augmentation techniques:
\begin{itemize}
    \item \textbf{Random rotation:} Each image slice is randomly rotated by an angle within $[-15^\circ, 15^\circ]$ to simulate different orientations.
    \item \textbf{Random flipping:} Horizontal and vertical flipping are applied with a probability of $0.5$ to introduce spatial variability.
    \item \textbf{Scaling:} The images are randomly scaled within the range $[0.9, 1.1]$ to enhance robustness to size variations.
    \item \textbf{Elastic deformation:} Spatially elastic transformations are applied to simulate realistic deformations in medical images.
    \item \textbf{Contrast and brightness adjustment:} To account for variations in image acquisition settings, we randomly adjust the contrast and brightness of images.
\end{itemize}
These augmentation strategies ensure that the model learns from diverse image distributions while preserving anatomical structures.

All preprocessing and augmentation operations are implemented using standard deep learning libraries, and applied online during training to maximize variability in training samples.

\section{Additional analysis}

\label{sec:ablation}
\subsection{Ablation study on the LoRA component}
We also conduct our additional ablation studies on 10\% Synapse dataset. In the Tab.~\ref{tab:rank}, we compare the effectiveness of the layers of LoRA component
among these models and their variants. From the result, we found that all models
and their variants, the best performance is achieved when the rank increases to 4,
while the performance drops when the rank increases to 16.

Furthermore, the model incorporating multiple LoRA experts exhibits a smaller
performance gap compared to the original model at different rank values, suggesting
that the proposed mechanisms enhance the model's robustness to variations in the
rank parameter.

\begin{table}[t]
   \caption{Effectiveness of HL-Attn compared to HCAT.}
   \label{tab:sup_effectiveness1}
   \vspace{-2.0em}
   \begin{center}
      \small
      \setlength{\tabcolsep}{5pt}
      \begin{tabular}{c| c c }
         \toprule \textbf{Model} & \textbf{Mean Dice(\%)} & \textbf{Inference Time(s)} \\
         \midrule HCAT           & \textbf{84.40 }        & 4.2                        \\
         HL-Attn (ours)          & 84.34                  & \textbf{3.1}               \\
         \bottomrule
      \end{tabular}
   \end{center}
   \vspace{-2.0em}
\end{table}

\begin{table*}[h]
\caption{\textbf{Full Effectiveness of Different Multiple LoRA experts Fusion Strategies}}
\label{tab:sup_effectiveness}
\vspace {-2.0em}
\begin{center}
\small
\setlength{\tabcolsep}{5pt}
\begin{tabular}{c c c c|c c c}
\toprule
\textbf{\makecell[c]{Image-level LoRA}} & \textbf{\makecell[c]{Patch-level LoRA}} & \textbf{\makecell[c]{Pixel-level LoRA}} & \textbf{\makecell[c]{Fusion Module}} & \textbf{\makecell[c]{10\% Synapse}} & \textbf{5\% LA} & \textbf{7.5\% PROMISE12}\\
\midrule
\cmark & \cmark & \cmark & LAC~\cite{lac1} & 82.41 & 90.01           & 88.97  \\
\cmark & \cmark & \cmark & MOLE~\cite{mole}& 83.91 & 91.59           & 89.44      \\
\cmark & \cmark & \cmark & LoRAHub~\cite{lorahub} & 81.07 & 88.31           & 87.43      \\
\cmark & \cmark & \cmark & HL-Attn (ours) & \textbf{84.34}  & \textbf{92.46}  & \textbf{90.14} \\
\midrule
1 & 1 & 2 & HL-Attn (ours) & 84.21 & 92.10 & 89.95\\
1 & 2 & 1 & HL-Attn (ours) & 83.86 & 91.80 & 89.72\\
2 & 1 & 1 & HL-Attn (ours) & \textbf{84.34} & \textbf{92.46} & \textbf{90.14}\\
\bottomrule
\end{tabular}
\end{center}
\vspace {-2.0em}
\end{table*}

\subsection{Additional study on the HL-Attn}

Our work focuses on proposing an innovative multi-level framework that
integrates existing methods in a novel way to address specific challenges in
medical image analysis. While we build upon widely recognized techniques like MAE
and SimCLRv2, leveraging strong foundations is common and necessary in advancing
research. The novelty of HL-Attn lies in the hierarchical design and effective combination
of these methods, with a focus on simplicity and adaptability. Even with a straightforward
fusion strategy, our approach demonstrates significant gains. To further validate
the effectiveness of our method, we conducted experiments on the hierarchical
cross-attention transformer (HCAT). The results (Tab.~\ref{tab:sup_effectiveness1}), demonstrate that HL-Attn
achieves comparable mean Dice scores with a reduction in inference time, highlighting
the efficiency of our framework.

\subsection{Effectiveness of Different Multiple LoRA experts Fusion Strategies}
We have further supplement our experiments on the LA and PROMISE12 datasets to provide
a more comprehensive assessment of SAMora's segmentation performance. The
results in Tab.~\ref{tab:sup_effectiveness} show that HL-Attn outperforms other fusion strategies across both
datasets, achieving the highest mean Dice scores. This demonstrates the effectiveness
of our proposed method in enhancing segmentation performance across different medical imaging tasks.

\subsection{Complementarity of multiple LoRAs}
\begin{figure*}[h]
   \centering
   \caption{Complementarity of multiple LoRAs.}
   \vspace{-1em}
   \includegraphics[width=1\linewidth]{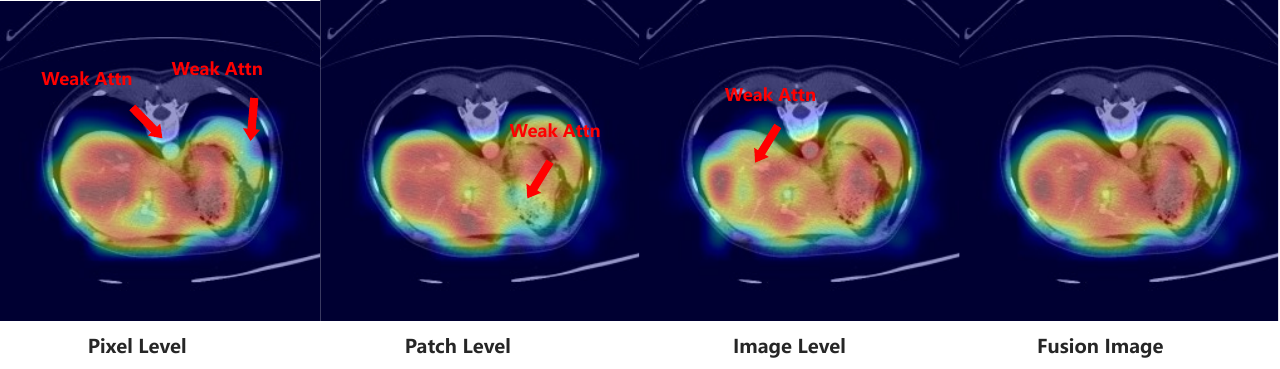}
   \label{fig:complement}
   \vspace{-2em}
\end{figure*}
 Furthermore,  the Fig.~\ref{fig:complement} illustrates the
  complementarity across the three LoRA levels. It shows
  that individual levels fail to capture certain structural details, while the fusion image effectively integrates these
  features, resulting in improved overall accuracy. This highlights how the
  hierarchical fusion leverages distinct strengths from each level. These visual results demonstrate that the modifications to the model architecture have successfully guided the network to concentrate on the most relevant features. 

\subsection{Clarification of Training Time}
The CPT process can be seen as an equivalent fine-tuning phase for SimCLRv2 and
MAE. For CPT, we sampled 100,000 images from datasets like AMOS and employed a
comprehensive pre-training process integrating SimCLRv2 and MAE to effectively
learn hierarchical features. As shown in the table, models with shorter CPT
durations demonstrate that SAMora can balance efficiency and performance. H-SAMora-T1,
which excludes CPT and performs minimal pre-training, achieves a Mean Dice of 80.72,
slightly outperforming H-SAM. H-SAMora-T2, with a reduced CPT duration of 0.8
hours, improves further to 80.97. The full CPT version, H-SAMora, achieves the
highest Mean Dice of 84.34, highlighting the benefits of a complete pre-training
process. These results confirm SAMora’s adaptability to different resource
constraints, as even shorter CPT durations deliver significant improvements,
while the default CPT duration maximizes performance and demonstrates the
framework’s full potential. The detailed training configurations and results will
be presented in the revised manuscript.

\begin{table}[h]
   \caption{Results of different training time of CPT}
   \vspace{-2.0em}
   \begin{center}
      \small
      \setlength{\tabcolsep}{3pt}
      \begin{tabular}{c|c c c| c}
         \toprule Model & CPT  & Pre-Training & Fine-Tuning & Mean Dice      \\
         \midrule H-SAM & -    & -            & 2           & 80.35          \\
         H-SAMora-T1    & -    & 1.8          & 0.1         & 80.72          \\
         H-SAMora-T2    & 0.8  & 1.3          & 0.1         & 80.97          \\
         H-SAMora       & 12.7 & 13.4         & 0.1         & \textbf{84.34} \\
         \bottomrule
      \end{tabular}
   \end{center}
   \label{tab:sup_sup_la_pr_results}
   \vspace{-2.0em}
\end{table}

\subsection{Statistical validation}
To address this concern, we conducted statistical validation to confirm the significance
of our performance improvements on the Synapse dataset. We performed a paired t-test
on the mean Dice scores of SAMora, SAMed, SAMora-2, SAMed-2, H-SAMora, and H-SAM.
The results show highly significant differences, such as H-SAMora versus H-SAM with
a p-value of $1.7 \times 10^{-8}$ and a 95\% confidence interval of [0.0329,
0.0407]. Similarly, SAMora outperforms SAMed with a p-value of 0.0167 and SAMora-2
outperforms SAMed-2 with a p-value of 0.0281. These statistical tests validate
the robustness and significance of the reported improvements, and the detailed analysis
will be included in the revised manuscript.

\begin{table*}[t]
\caption{
\textbf{Full Performance Comparison of SAM and SAM2 Variants on Synapse Dataset. }
\textbf{Bold} numbers indicate the best performance.
By default, we utilize SAM as our base model.
$^{\dagger}$ indicates H-SAM based model; 
$^{\ast}$ indicates SAM2 based model.
}
\begin{center}
\vspace {-1.0em}
\small
\setlength{\tabcolsep}{3pt}
\begin{tabular}{c|c|c c c c c c c c|c c}
\toprule
\textbf{\makecell[c]{Training\\Set}} & \textbf{Method} & \textbf{Spleen} & \textbf{\makecell[c]{Right\\Kidney}} & \textbf{\makecell[c]{Left\\Kidney} } & \textbf{Gallbladder} & \textbf{Liver} & \textbf{Stomach} & \textbf{Aorta} & \textbf{Pancreas} & \textbf{Mean Dice}  $\textcolor{red}{\uparrow}$ & \textbf{HD} $\textcolor{darkgreen}{\downarrow}$\\
\midrule
\multirow{8}*{10\%} & AutoSAM~\cite{autosam}  & 68.80 & 77.44 & 76.53 & 24.87 & 88.06 & 52.70 & 75.19 & 34.58 & 55.69 & 31.67 \\
~ & SAM Adapter~\cite{samad} & 72.42 & 68.38 & 66.77 & 22.38 & 89.69 & 53.15 & 66.74 & 26.76 & 58.28 & 54.22 \\
~ & SAMed~\cite{samed}  & 85.82 & 82.25 & 82.62 & 63.15 & 92.72 & 67.20 & 78.72 & 52.12 & 75.57 & 23.02 \\
~ & SAMora (Ours) & \textbf{88.04} & \textbf{83.41} &\textbf{ 86.07} &\textbf{ 67.33} & \textbf{94.27} & \textbf{69.20} & \textbf{82.85} & \textbf{64.13} & \textbf{79.41} & \textbf{15.68} \\
\cmidrule(lr){2-12} 
~ & SAMed-2$^{\star}$ & 86.61 & 83.01 & 84.56 & 61.51 & 91.07 & 69.02 & 77.99 & 52.09 & 76.68 & 18.93 \\
~ & SAMora-2$^{\star}$ (Ours) & \textbf{87.81} & \textbf{85.73} & \textbf{86.35} & \textbf{68.30} & \textbf{93.78} & \textbf{75.24} & \textbf{81.12} & \textbf{63.62} & \textbf{80.24} & \textbf{16.27}\\
\cmidrule(lr){2-12} 
~ & H-SAM~\cite{hsam} & 90.21 & 84.16 & 85.65 & 70.70 & 94.29 & 76.10 & 85.54 & 56.17 & 80.35 & 15.54 \\
~ & H-SAMora$^{\dagger}$ (Ours) & \textbf{92.46} & \textbf{85.13} & \textbf{86.71} & \textbf{73.15} & \textbf{95.82} & \textbf{81.85} & \textbf{88.56} & \textbf{72.72} & \textbf{84.34} & \textbf{11.63}\\
\midrule        
\multirow{17}*{\makecell[c]{Fully\\Supervised}}  & TransUNet~\cite{transunet} & 81.87& 85.08&77.02& 63.16& 94.08&75.62& 87.23& 55.86& 77.49& 31.69 \\
~ & UNETR~\cite{unetr} & 85.60 & 85.00&  84.52&  56.30&  94.57&  70.46&  89.80&  60.47 & 78.35 &  18.59 \\
~ & SwinUnet~\cite{Swin-unet} & 85.47 & 66.53 & 83.28 & 79.61 & 94.29 & 56.58 & 90.66 & 76.60 & 79.13& 21.55 \\
~ & TransDeepLab~\cite{deeplab} & 86.04 & 69.16 & 84.08 & 79.88 & 93.53 & 61.19 & 89.00 & 78.40 & 80.16 & 21.25 \\
~ & DAE-Former~\cite{dae} & 88.96 & 72.30 & 86.08 & \textbf{80.88} & 94.98 & 65.12 & 91.94 & 79.19 & 82.43 & 17.46 \\
~ & MERIT~\cite{merit} & \textbf{92.01} & 84.85 & \textbf{87.79} & 74.40 & 95.26 & 85.38 & 87.71 & 71.81 & 84.90 &  13.22 \\
~ & nnFormer~\cite{nnformer} & 86.57 & 90.51 &86.25 &70.17 &\textbf{96.84} & \textbf{86.83} & 92.04 & \textbf{83.35} & 86.57 &   10.63 \\
~ & UNETR++~\cite{unetr++} & 87.54& \textbf{95.77}&  87.18&  71.25& 96.42& 86.01& \textbf{ 92.52}& 81.10 &  \textbf{87.22} & \textbf{ 7.53} \\
\cmidrule(lr){2-12} 
~ & SAM Adapter~\cite{samad} & 83.68 & \textbf{79.00} & 79.02 & 57.49 & 92.67 & 69.48 & 77.93 & 43.07 & 72.80 & 33.08 \\
~ & SAM3D~\cite{sam3d} &  84.29 &  85.64 & \textbf{86.31} &  49.81 &  95.42 & \textbf{76.11} &  \textbf{89.57} & 69.32 &  79.56 &  17.87 \\
~ & SAMed~\cite{samed}  & 87.77 & 69.11 & 80.45 & 79.95 & 94.80 & 72.17 & 88.72 & 82.06 & 81.88 & 20.64 \\
~ & SAMora (Ours) & \textbf{89.27} & 74.05 & \textbf{81.04} & \textbf{81.51} & \textbf{94.97} & 74.53 & 88.87 & \textbf{82.42} & \textbf{83.33} & \textbf{14.57} \\

\cmidrule(lr){2-12} 
~ & SAMed-2$^{\star}$ & 88.63 & 68.63 & 81.22 & 80.33 & 95.18 & 71.00 & 87.63 & 81.93 & 82.12 & 12.76 \\
~ & SAMora-2$^{\star}$ (Ours) & \textbf{91.78} & \textbf{75.85} & \textbf{82.02} & \textbf{83.52} & \textbf{95.49} & \textbf{75.11} & \textbf{87.11} & \textbf{82.26} & \textbf{84.14} & \textbf{10.28}\\
\cmidrule(lr){2-12} 
~ & H-SAM~\cite{hsam} & 93.34 & 89.93 & 91.88 & 73.49 & 95.72 & 87.10 & 89.38 & 71.11 & 86.49 & 8.18 \\
~ & H-SAMora$^{\dagger}$ (Ours) & \textbf{94.62} &\textbf{ 91.45} & \textbf{93.00} & \textbf{76.55} & \textbf{96.51} &\textbf{ 89.95} &\textbf{ 89.55} & \textbf{77.09} & \textbf{88.59} & \textbf{7.09}\\
\bottomrule
\end{tabular}
\label{tab:sup_sam_comparison}
\end{center}
\vspace {-2.0em}
\end{table*}
Although SAMora performs well in most medical image segmentation tasks, its
performance may degrade when handling noisy or low-quality images. Future research
could focus on improving the model’s robustness to such challenging image quality
issues.

\subsection{Complete Experimental Results}
This section presents the full experimental results only partially included in the main text, providing a more comprehensive evaluation of the proposed method. Table 13 offers a detailed performance comparison of SAM and SAM2 variants on the Synapse dataset, where bold numbers indicate the best performance. Table 14 extends the comparison by benchmarking various SAM variants against multiple semi-supervised methods across different datasets. Additionally, Table 15 provides a complete ablation analysis of multiple LoRA experts on the 10\% Synapse dataset, where "Scratch" refers to models trained from scratch. At the same time, "T-S" denotes training using a teacher-student framework. These tables collectively reinforce the findings and conclusions drawn in the main text, offering more profound insights into the effectiveness of the proposed approach.

\begin{table}[t]
\caption{\textbf{Full Comparison of SAM Variants against Semi-Supervised Methods across Various Datasets}}
\vspace {-2.0em}
\begin{center}
\small
\setlength{\tabcolsep}{7pt}
\begin{tabular}{c|c|c|c}
\toprule
\textbf{Method} &\textbf{\makecell[c]{10\% \\Synapse}} & \textbf{\makecell[c]{5\%\\ LA}}  & \textbf{\makecell[c]{7.5\% \\PROMISE12}}  \\
\midrule
nnUnet~\cite{nnUnet}        & - & 64.02     & 84.22      \\
UA-MT~\cite{uamt}        &  - & 82.26    & 65.05      \\
SS-Net~\cite{ssnet}        & 56.74 & 86.33    & 73.19      \\
MC-Net~\cite{mcnet}        & 61.20 & 83.59     & 72.66      \\
DTC~\cite{dtc}        & - & 81.25     & 63.44      \\
\midrule
AutoSAM~\cite{autosam}        & 55.69 & 74.73     & 68.40      \\
SAM Adapter~\cite{samad}      & 58.28 & 82.79     & 75.45    \\
SAMed~\cite{samed}               & 75.57 & 87.72     & 86.00   \\
SAMora (Ours)               & \textbf{79.41} & \textbf{90.13}  & \textbf{88.44}   \\
\midrule
SAMed-2              & 76.68  & 87.91     & 86.50      \\
SAMora-2 (Ours)             & \textbf{80.24} & \textbf{91.04}  & \textbf{89.27}  \\
\midrule
H-SAM~\cite{hsam}               & 80.35 & 89.22     & 87.27      \\
H-SAMora (Ours)             & \textbf{84.34} & \textbf{92.46}  & \textbf{90.14}  \\
\bottomrule
\end{tabular}
\end{center}
\label{tab:sup_la_pr_results}
\vspace {-2.0em}
\end{table}


\begin{table}[t]
\caption{\textbf{Full Ablation Analysis of Multiple LoRA experts on 10\% Synapse.} ``Scratch'' means the model is trained from scratch, while ``T-S'' indicates the model is trained by the Teacher-Student framework}
\label{tab:sup_ablation}
\vspace {-2.0em}
\begin{center}
\small
\setlength{\tabcolsep}{0.5pt}
\begin{tabular}{c c c c|c}
\toprule
\textbf{\makecell[c]{Image-level\\LoRA}} & \textbf{\makecell[c]{Patch-level\\LoRA}}  & \textbf{\makecell[c]{Pixel-level\\LoRA}} & \textbf{Model} & \textbf{\makecell[c]{Mean Dice\\(\%)}} \\
\midrule
\textbf{Scratch}  & \xmark  & \xmark &SAMora & 77.20 \\
\textbf{T-S} ($w/o$ CPT) & \xmark & \xmark & SAMora & 77.31 \\
\textbf{T-S} ($w/$ CPT) & \xmark  & \xmark &SAMora & \textbf{78.03} \\
\textbf{Scratch}  & \xmark  & \xmark &H-SAMora & 82.09 \\
\textbf{T-S} ($w/o$ CPT) & \xmark & \xmark & H-SAMora & 82.17 \\
\textbf{T-S} ($w/$ CPT) & \xmark  & \xmark &H-SAMora & \textbf{82.65} \\
\midrule
\xmark  &  \textbf{Scratch} & \xmark & SAMora & 76.54 \\
\xmark  &  \textbf{T-S} ($w/o$ CPT) & \xmark & SAMora & 77.19 \\
\xmark  &  \textbf{T-S} ($w/$ CPT) &  \xmark &SAMora & \textbf{78.81} \\
\xmark  &  \textbf{Scratch} & \xmark & H-SAMora & 81.67 \\
\xmark  &  \textbf{T-S} ($w/o$ CPT) & \xmark & H-SAMora & 82.04 \\
\xmark  &  \textbf{T-S} ($w/$ CPT) &  \xmark &H-SAMora & \textbf{83.02} \\
\midrule
\xmark & \xmark &  \textbf{Scratch}  & SAMora & 76.97 \\
\xmark & \xmark &  \textbf{Scratch}  & H-SAMora & 81.58 \\
\bottomrule
\end{tabular}
\end{center}
\vspace {-2.5em}
\end{table}

\section{Limitation and Future Work}
\label{supsec:limit} Despite the promising results of SAMora, several
limitations need to be addressed in future research.

While SAMora reduces the reliance on labeled data through self-supervised
learning, it still requires some labeled data for fine-tuning. Therefore, further
exploration of fully unsupervised data is needed. On the other hand, we observe that
weakly labeled data, compared to fully labeled data, has been widely applied in research
due to its lower cost and reduced need for manual annotation, which makes it
more scalable and practical in real-world applications~\cite{part1,part2}. Consequently,
future work will explore integrating weakly labeled data to enhance SAMora’s
performance, allowing it to better generalize across a broader range of medical image
segmentation tasks.


\end{document}